\documentclass[hyperref,colorlinks=true, allcolors=blue,journal=jctcce,manuscript=article]{achemso}
\setkeys{acs}{doi=true}
\usepackage[english]{babel}

\usepackage[utf8]{inputenc} 
\usepackage{amsmath}
\usepackage{amsfonts}
\usepackage{amssymb}
\usepackage{mathtools}
\usepackage{graphicx}
\usepackage{physics}
\usepackage{cancel}
\usepackage{tikz}
\usepackage{circuitikz}
\usetikzlibrary{fit}
\usetikzlibrary{positioning}
\usetikzlibrary{decorations.pathmorphing, decorations.pathreplacing, decorations.markings, backgrounds}
\usepackage{tikz-network}
\usepackage[justification=centering]{caption,subcaption}
\usepackage{makecell}
\usepackage{afterpage}
\usepackage[ruled, vlined, linesnumbered]{algorithm2e}
\usepackage{colortbl}
\usepackage{xcolor
}
\usepackage{setspace}
\usepackage[version=4]{mhchem}
\usepackage[nohyperlinks, nolist]{acronym}
\usepackage{mciteplus}

\usetikzlibrary{arrows.meta,positioning,shapes.geometric,calc}

\definecolor{swordblue}{RGB}{36,74,210}
\definecolor{lightblue}{RGB}{235,241,255}
\definecolor{softgray}{RGB}{246,246,246}
\definecolor{successgreen}{RGB}{225,246,231}
\definecolor{decisionyellow}{RGB}{255,247,214}

\tikzset{
  flowarrow/.style={
    -{Stealth[length=3.2mm,width=2.2mm]},
    line width=1.15pt,
    draw=swordblue,
    rounded corners=4pt
  },
  process/.style={
    rectangle,
    rounded corners=3pt,
    draw=swordblue,
    line width=1.15pt,
    fill=lightblue,
    text width=10.6cm,
    minimum height=1.35cm,
    align=center,
    inner sep=7pt,
    font=\large
  },
  state/.style={
    rectangle,
    rounded corners=3pt,
    draw=black!70,
    line width=1.15pt,
    fill=softgray,
    text width=10.6cm,
    minimum height=1.25cm,
    align=center,
    inner sep=7pt,
    font=\large
  },
  decision/.style={
    diamond,
    aspect=2.3,
    draw=black!75,
    line width=1.15pt,
    fill=decisionyellow,
    text width=5.4cm,
    align=center,
    inner sep=3pt,
    font=\large
  },
  terminal/.style={
    rectangle,
    rounded corners=6pt,
    draw=green!45!black,
    line width=1.3pt,
    fill=successgreen,
    minimum width=4.2cm,
    minimum height=1.15cm,
    align=center,
    font=\bfseries\Large
  },
  smalllabel/.style={
    fill=white,
    inner sep=2pt,
    font=\normalsize
  }
}

\graphicspath{{figures/}}
\makeatletter
\def\input@path{{figures/}}
\makeatother

\pgfdeclaredecoration{simple line}{start}
{
  \state{start}[width = +0pt,
                next state=step]{
    \pgfpathmoveto{\pgfpoint{0pt}{0pt}}
  }
  \state{step}[auto end on length    = 3pt,
               auto corner on length = 3pt,               
               width=+1pt]
  {
    \pgfpathlineto{\pgfpoint{1pt}{0pt}}
  }
  \state{final}
  {}
}

\makeatletter
\def\pgf@stroke@inner@line@if@needed{%
  \ifdim\pgfinnerlinewidth>0pt\relax%
    \let\pgf@temp@save=\pgf@strokecolor@global
    \pgfsys@beginscope%
    {%
      \pgfsetrectcap
      \pgfsys@setlinewidth{\pgfinnerlinewidth}%
      \pgfsetstrokecolor{\pgfinnerstrokecolor}%
      \pgfsyssoftpath@invokecurrentpath%
      \pgfsys@stroke%
    }%
    \pgfsys@endscope%
    \global\let\pgf@strokecolor@global=\pgf@temp@save
  \fi%
}

\title{
Spinorial Weyl Operator Reduction of Dirac (SWORD): Exact Relativistic Quantum Chemistry with Half the Spinor }

\author{Jacopo Masotti}
\email{jacopo.masotti@uit.no}
\author{Quentin Pitteloud}
\email{quentin.pitteloud@uit.no}
\author{Luca Frediani}
\email{luca.frediani@uit.no}
\affiliation{Hylleraas Centre, Department of Chemistry, UiT University of Troms\o{}, The Arctic University of Norway, N-9037 Troms\o{}, Norway}

\newlength{\dch}
\newlength{\sch}

\newcommand{\scaling}{\phi}
\newcommand{\wavelet}{\psi}

\newcommand{\major}{\cellcolor{blue!100}}
\newcommand{\minor}{\cellcolor{blue!25}}

\newcommand{\vampyr}{\texttt{VAMPyR}\cite{bjorgve_vampyr_2024}}
\newcommand{\remrchem}{\texttt{ReMRChem}\cite{ReMRChem}}
\newcommand{\mrchem}{\texttt{MRChem}\cite{Wind_2023}}
\newcommand{\mrcpp}{\texttt{mrcpp}\cite{bast_2023_7967323}}
\newcommand{\grasp}{\texttt{GRASP}\cite{grasp_ref}}

\newcommand{\orbright}{\ensuremath{\psi^R}}
\newcommand{\orbleft}{\ensuremath{\psi^L}}


\begin{document}
\begin{acronym}
\acro{ABGV}{Alpert, Beylkin, Ginez and Vozovoi}
\acro{ae}{almost everywhere}
\acro{AO}{atomic orbital}
\acro{API}{Application Programmer Interface}
\acro{AUS}{Advanced User Support}
\acro{BEM}{Boundary Element Method}
\acro{BO}{Born-Oppenheimer}
\acro{BS}{B-Spline}
\acro{BVP}{boundary value problem}
\acro{CBS}{complete basis set}
\acro{CC}{Coupled Cluster}
\acro{CoE}{Centre of Excellence}
\acro{CTCC}{Centre for Theoretical and Computational Chemistry}
\acro{DC}{dielectric continuum}
\acro{DD}{domain decomposition}
\acro{DCHF}{Dirac-Coulomb Hartree-Fock}
\acro{DFT}{density functional theory}
\acro{DHF}{Dirac--Hartree--Fock}
\acro{DIIS}{Direct Inversion in the Iterative Subspace}
\acro{DKH}{Douglas-Kroll-Hess}
\acro{ECP}{effective core potential}
\acro{EFP}{effective fragment potential}
\acro{EU}{European Union}
\acro{FMM}{fast multipole method}
\acro{GGA}{generalized gradient approximation}
\acro{GPE}{Generalized Poisson Equation}
\acro{GTO}{Gaussian Type Orbital}
\acro{HF}{Hartree-Fock}
\acro{HPC}{high-performance computing}
\acro{Hylleraas}[HC]{Hylleraas Centre for Quantum Molecular Sciences}
\acro{IEF}{Integral Equation Formalism}
\acro{IEFPCM}{Integral Equation Formalism \ac{PCM}}
\acro{IGLO}{individual gauge for localized orbitals}
\acro{KAIN}{Krylov-accelerated inexact Newton}
\acro{KB}{kinetic balance}
\acro{KS}{Kohn-Sham}
\acro{KTRS}{Kramers' Time Reversal Symmetry}
\acro{LAO}{London atomic orbital}
\acro{LAPW}{linearized augmented plane wave}
\acro{LDA}{local density approximation}
\acro{MAD}{mean absolute deviation}
\acro{maxAD}{maximum absolute deviation}
\acro{MCSCF}{multiconfiguration self consistent field}
\acro{MM}{molecular mechanics}
\acro{MO}{molecular orbital}
\acro{MPA}{multiphoton absorption}
\acro{MRA}{multiresolution analysis}
\acro{MSDD}{Minnesota Solvent Descriptor Database}
\acro{MW}{multiwavelet}
\acro{NAO}{numerical atomic orbital}
\acro{NeIC}{nordic e-infrastructure collaboration}
\acro{NMR}{nuclear magnetic resonance}
\acro{NP}{nanoparticle}
\acro{NS}{non-standard}
\acro{OLED}{organic light emitting diode}
\acro{PAW}{projector augmented wave}
\acro{PBC}{Periodic Boundary Condition}
\acro{PCM}{polarizable continuum model}
\acro{PDE}{partial differential equation}
\acro{PW}{plane wave}
\acro{QC}{quantum chemistry}
\acro{QM}{quantum mechanics}
\acro{QM/MM}{quantum mechanics/molecular mechanics}
\acro{RCN}{Research Council of Norway}
\acro{RDM}{1-body reduced density matrix}
\acro{RI}{resolution of the identity}
\acro{RKB}{restricted kinetic balance}
\acro{RMSD}{root mean square deviation}
\acro{SAPT}{symmetry-adapted perturbation theory}
\acro{SC}{semiconductor}
\acro{SCF}{self-consistent field}
\acro{SCRF}{self-consistent reaction field}
\acro{SERS}{surface-enhanced raman scattering}
\acro{SI}{Supporting Information}
\acro{STSM}{short-term scientific mission}
\acro{TRS}{time-reversal symmetry}
\acro{WP}{Work Package}
\acro{WPREL}[WP1]{Work Package 1}
\acro{WPROP}[WP2]{Work Package 2}
\acro{WPAPP}[WP3]{Work Package 3}
\acro{X2C}{exact two-component}
\acro{ZORA}{zero-order relativistic approximation}
\end{acronym}

\begin{abstract}
We present an exact self-consistent two-component reformulation of the Dirac equation in the Weyl representation. By exploiting the diagonal structure of the kinetic operator and the purely scalar mass coupling in the Weyl basis, the four-component Dirac problem is reduced to an exact two-component form without introducing inverse-potential operators that arise in conventional energy-decoupling approaches. The resulting second-order differential equations are cast into an integral formulation and solved iteratively via convolution with the Helmholtz Green's function. We implement this formalism within an adaptive multiwavelet basis framework, providing rigorous, user-defined error control. Proof-of-concept calculations for simple atomic systems demonstrate the numerical stability, accuracy, and efficiency of the proposed algorithm. 
\end{abstract}

\maketitle

\section{Introduction}
Relativistic effects are indispensable components of modern electronic structure theory, governing a wide array of chemical and physical phenomena in systems containing heavy elements\cite{Das2023Relativistic}. Rather than serving as minor perturbative corrections, these effects dictate technologically relevant properties in materials and molecules, as extensively verified by high-resolution spectroscopy\cite{Ehrman2024Unveiling}. Consequently, the development of robust and accurate relativistic electronic Hamiltonians remains a central objective in quantum chemistry.

The introduction of the Dirac equation in 1928 provided the first fully relativistic description of the electron \cite{Original_Dirac}. However, it introduced a four-component wavefunction whose computational treatment poses significant challenges due to increased degrees of freedom and the presence of negative-energy states\cite{VAR_Collapse}. Geometrically, the Dirac equation demonstrates that fermions such as electrons are fundamentally described by spinors \cite{lounesto_clifford_2001}. Spinors are algebraic and geometric objects embedded within Clifford algebras, and while elegant, their abstract nature complicates direct numerical implementations\cite{doran_geometric_2003}. To make these objects computationally tractable, Dirac embedded the algebraic relations of spinors into a concrete matrix representation by mapping the four basis vectors of spacetime to the $4 \times 4$ complex gamma matrices $\gamma_\mu$.

Since physical observables are invariant under the choice of representation, the computational efficiency of a relativistic formulation may depend more strongly on the resulting operator structure than on the underlying physics itself. This motivates investigating whether alternative representations expose algebraic structures that are hidden in the conventional Dirac basis and can be exploited numerically.
From the mathematical point of view, an infinite number of matrix representations satisfy the underlying Clifford algebra required by the Dirac equation, each yielding a mathematically distinct but physically equivalent Hamiltonian.

The standard formulation of the Dirac equation in terms of large and small components has several advantages in connection with a \ac{AO} representation: it naturally yields the non relativistic limit (small components vanish and large components give electrons with different spin). Furthermore, it exploits the kinetic balance to link the basis set for the small components to the large one providing a clear path towards fewer component methods which assume that the small components can be treated adequately through a series of approximations.

Over the past five decades, a primary goal in relativistic quantum chemistry has therefore been to attempt to retain the rigorous accuracy of the four-component Dirac equation while reducing its effective dimensionality \cite{dyall_relativistic_2007}. Recognizing that chemical phenomena are dominated by positive-energy electronic states, various decoupling strategies have been developed, including the Foldy-Wouthuysen (FW), Douglas--Kroll--Hess (DKH), Barysz--Sadlej--Snijders (BSS), and exact two-component (X2C) methods \cite{Primer_Trond}. While successful, these standard energy-decoupling procedures typically yield highly complex operators involving inverse potential terms of the form $\propto 1/\hat{V}$. These terms become ill-defined or computationally intractable when non-local operators, such as the Hartree--Fock exchange operator $\hat{K}$, are present.

A transformation to the chiral representation, also known as the Weyl basis, allows the Dirac problem to be reformulated as an exact, self-consistent two-component equation, completely eliminating the need to explicitly propagate all four spinor components within the SCF procedure. In contrast to the large/small separation, which requires eventually an approximation, the Weyl representation is formally exact: the differential kinetic operator acting on the Weyl components is strictly diagonal, while the off-diagonal block is a simple scalar, in the form of the mass-coupling term ($mc^2$). This structure enables an exact reformulation using only a single chiral component, bypassing the singular $1/\hat{V}$ terms characteristic of conventional energy-based decouplings.

To the best of our knowledge, the Weyl representation has only been explored computationally for one-electron problems\cite{Karwowski101002qua21053, STANKE1010071-4020-4528-x_10}, underlying the difficulty/inertia in extending such an approach to the general molecular case. This can be traced to the difference of the Left and Right chiral components from standard non-relativistic atomic orbitals, the underlying computational basis must be highly flexible and capable of executing integral operators efficiently. Multiwavelets provide an ideal framework for this task, having been designed specifically for the adaptive solution of differential and integral equations with rigorous, user-specified error control\cite{harrison_madness_2004}. 
Furthermore, manipulating the Dirac Hamiltonian in the chiral basis reveals a second-order structure that naturally lends itself to an integral Green's-function formulation solved via iterative convolutions\cite{Frediani2013_Helmholtz_poissan,Jensen2022_integral_formalism}. 

Here, we combine this Weyl-based exact two-component reformulation with the adaptive multiwavelet framework implemented in the \vampyr{} software library. Proof-of-concept calculations demonstrate the viability of this approach and establish a foundational framework for future relativistic electronic structure methods. A similar approach has already been employed with MW at the scalar and 4-component level \cite{Scalar_and_ZORA_MW,masotti2026}. This work extends these previous developments.


\section{Theoretical Background}
\subsection{The Dirac Equation in Clifford Algebra}
The computational form of the Dirac equation depends on the chosen representation, although the physical predictions do not. This observation motivates the search for representations that expose computationally advantageous structures.
In its original derivation, the Dirac equation can be motivated by factoring the Klein--Gordon operator:
\begin{equation}
    -(\square +m^2c^2)=(i\gamma^\mu \partial_\mu -mc)(i\gamma^\nu \partial_\nu +mc)
\end{equation}
where $\square$ is the d'Alembert operator, under the condition that the coefficients $\gamma^\mu$ satisfy the Clifford algebra anticommutation relations:
\begin{equation}\label{eq:gamma_relations}
    \qty{\gamma^\mu,\gamma^\nu}= 2\eta^{\mu\nu}\mathbb{I}_2,
\end{equation}
with the Minkowski metric tensor chosen as $\eta^{\mu\nu}= \text{diag}(+1,-1,-1,-1)$ and where $\mathbb{I}_4$ denotes the $4 \times 4$ identity matrix. Identifying the time derivative with the energy operator via $\hat{H} = i \partial_t=ic\partial_{x^0}$, we isolate the time evolution term to obtain:
\begin{equation}
    i\,\gamma^0 \pdv{t} \Psi= -ic\gamma^j\partial_{j}\Psi +mc^2\Psi.
\end{equation}
Multiplying from the left by $\gamma^0$, utilizing the property $(\gamma^0)^2=\mathbb{I}_4$, and defining the standard matrices $\alpha^j = \gamma^0\gamma^j$ and $\beta = \gamma^0$, we arrive at the standard free Dirac Hamiltonian:
\begin{equation}
    \hat{h}_D= c (\Vec{\alpha}\cdot \Vec{p}) + \beta \,mc^2.
\end{equation}
This expression is independent of the particular matrix representation chosen for the $\gamma^\mu$ and follows directly from the Clifford algebra relations in Eq.~\eqref{eq:gamma_relations}, with the usual RQC matrices $\alpha^j$ and $\beta$ expressed in terms of the gamma matrices.

\subsubsection{Spinor Basis Representations}
The algebraic relations governing the $\gamma$ matrices define a specific representation of the Clifford product and its basis elements. Universally, a Clifford algebra $\text{Cl}(V,g)$ defined over a vector space $V$ with a metric $g$ can be constructed as the quotient of the tensor algebra $T(V)$ with respect to a two-sided ideal $\mathcal{I}$ generated by the metric relations:
\begin{equation}
    \text{Cl}(V,g) = T(V)/\mathcal{I}.
\end{equation}
This algebraic structure does not dictate a unique matrix representation; hence, infinitely many equivalent matrix representations exist that embed the complexified spacetime algebra $\mathbb{C}\otimes \text{Cl}_{1,3}$ into matrix algebra\cite{Cliff_Alg_Introduction_book}.

In relativistic quantum chemistry, it is standard practice to employ $4\times4$ complex matrices, yielding a $4\times4$ matrix differential operator for the Hamiltonian. While alternative choices involving real numbers ($\mathbb{R}$) or quaternions ($\mathbb{H}$) can alter the number of components within the representation, the underlying physical meaning remains invariant\cite{lounesto_clifford_2001}. Changing the representation merely alters the explicit form of the differential equations and their resulting solutions.

Geometrically, a four-component Dirac spinor $\Psi\in \mathbb{C}\otimes Cl(1,3)$ can always be decomposed into two two-component Weyl spinors, a property rooted in the structure of the complexified spacetime algebra \cite{Cliff_Alg_Introduction_book, doran_geometric_2003, lounesto_clifford_2001}. However, the physical interpretation of these subcomponents depends on the chosen representation.

This allows us to find any arbitrary choice of matrix to represent $\gamma^\mu$, as long as they obey the correct anti-commutation rules enforced by the Clifford algebra. Every different representation choice will change the way the Dirac Hamiltonian looks and thus the solution spinor. At the same time the meaning one assigns to each of these components in the Dirac spinor, depends on the matrix representation chosen such that the underlying meaning of this object will not change. 

\paragraph{The Dirac Representation}
The standard choice in relativistic quantum chemistry is the Dirac representation:
\begin{equation}\label{eq.dirac_rep}
    \gamma^0 = \begin{pmatrix}
        \mathbb{I}_2 & 0\\ 0 & -\mathbb{I}_2
    \end{pmatrix}, \quad \gamma^i = \begin{pmatrix}
        0 & \sigma^i \\-\sigma^i &0
    \end{pmatrix}, \quad \Psi = (\psi^\alpha_L,\ \psi^\beta_L, \ \psi^\alpha_S, \ \psi^\beta_S)^T
\end{equation}
Here, the four-component spinor is the direct sum of the "Large" ($\psi_L$) and "Small" ($\psi_S$) two-spinor components. This terminology reflects their relative magnitudes for positive-energy electronic states in the non-relativistic limit, where the small component is suppressed relative to the large component by a factor of order $v/c$. The diagonal form of $\gamma^0$ makes the rest-mass contribution block diagonal, rendering this representation particularly convenient for the description and separation of electronic states in relativistic quantum chemistry.

\paragraph{The Weyl Representation}
The chiral or Weyl representation is defined by choosing an off-diagonal $\gamma^0$:
\begin{equation}\label{eq:Weyl_representation}
    \gamma^0 = \begin{pmatrix}
        0&\mathbb{I}_2 \\ \mathbb{I}_2 & 0
    \end{pmatrix}, \quad \gamma^i = \begin{pmatrix}
        0 & \sigma^i \\-\sigma^i &0
    \end{pmatrix},\quad \Psi = (\psi^\alpha_L,\ \psi^\beta_L, \ \psi^\alpha_R, \ \psi^\beta_R)^T
\end{equation}
In this basis, the Dirac spinor decomposes into Left-handed ($\psi_L$) and Right-handed ($\psi_R$) chiral components, since both Large and Left start with the same letter, to avoid confusion we'll only refer to the Left component from now on unless stated differently. The resulting Dirac Hamiltonian takes the block form:
\begin{equation}\label{eq.Weyl_Hamiltonisn}
    \hat{H}_W = \begin{pmatrix}
        -c(\vec\sigma\cdot \hat{p})+\hat V \, \mathbb{I}_2& mc^2\, \mathbb{I}_2\\
        mc^2\, \mathbb{I}_2 & +c(\vec{\sigma} \cdot \hat p )+\hat V\, \mathbb{I}_2
    \end{pmatrix},
\end{equation}
where the differential operators are confined to the diagonal blocks, and the off-diagonal coupling is a simple scalar mass term.
This separation of differential and mass operators is the key structural property exploited throughout the remainder of this work.

Although these two representations are connected by a transformation, this should not be confused with a rotation. In fact the operation that allows to go from Weyl to Dirac and \textit{vice-versa} is a \emph{similarity transform}.
Given any Dirac spinor in any representation, one can always extract the chiral components via the projector $P_{L/R} =\frac{1}{2}(1\mp i \gamma^0\gamma^1\gamma^2\gamma^3)$. When it comes to the chiral representation using the gamma matrices showed in \eqref{eq:Weyl_representation}, the $P_{L/R}$ projector is block diagonal.

Such modified Hamiltonian still share the same exact spectrum of the Dirac representation one, but the solutions to it are quite different in properties. The most notable fact is that the 2-spinors composing the solution are not eigenstates of $\hat{L}^2$ and thus can not be addressed with the usual labels of symmetry such as $s$, $p$, $d$ or $f$. This implies that the wavefunction of the Right and Left spinor cannot be easily factorized in a radial part times an angular one, even in the single electron case \cite{Karwowski101002qua21053}. This could explain why no relevant atomic/molecular have been published in the past years, even if this representation has long been studied and understood in other fields. 

For this reason MW are the natural choice to represent such odd functions, not relying on any function factorization moreover expressing the function in an adaptive way.

\section{Non-relativistic limit in the two representations and kinetic balance}

As discussed above, changing the representation of the $\gamma^\mu$ matrices changes the explicit form of the four-component spinor without altering its physical content. This becomes particularly apparent in the non-relativistic limit and helps explain why the Dirac representation has traditionally been preferred in relativistic quantum chemistry. It also has a practical consequence for the present work, since readily available initial guesses are generally obtained from a non-relativistic electronic-structure calculation and must therefore be embedded into the relativistic spinor space.

Let $\Phi$ denote a two-component non-relativistic Pauli spinor. For example, for a spin-up orbital one may write $\Phi=(\varphi,0)^T$, where $\varphi$ is the corresponding scalar spatial orbital. In the Dirac representation, a positive-energy solution is partitioned into large and small components,
\begin{equation}
\Psi_D =
\begin{pmatrix}
\psi_L^{D} \\
\psi_S^{D}
\end{pmatrix}.
\end{equation}
After separating the total relativistic energy as
\begin{equation}
E = mc^2+\varepsilon,
\end{equation}
the lower block of the Dirac equation gives
\begin{equation}
\psi_S^{D} = \frac{c{\vec\sigma}\cdot\hat{ p}} {E-\hat V+mc^2}, \psi_L^{D}.
\end{equation}
For positive-energy states, taking the non-relativistic limit $c\rightarrow\infty$ therefore yields
\begin{equation}
\label{eq:dirac_kinetic_balance}
\psi_S^{D} = \frac{{\vec \sigma}\cdot\hat{ p}}{2mc} \psi_L^{D} +\mathcal{O}(c^{-2}).
\end{equation}
Consequently, to zeroth order in $1/c$,
\begin{equation}
\Psi_D^{\mathrm{NR}} =
\begin{pmatrix}
\Phi\\
0
\end{pmatrix},
\end{equation}
while retaining the leading relativistic coupling gives the kinetically balanced spinor
\begin{equation}
\label{eq:dirac_balanced_guess}
\Psi_D^{\mathrm{KB}} =
\begin{pmatrix}
\Phi\\
\dfrac{\boldsymbol{\sigma}\cdot\hat{\mathbf p}}{2mc}\Phi
\end{pmatrix}
+\mathcal{O}(c^{-2}).
\end{equation}
Equation~\eqref{eq:dirac_kinetic_balance} is the usual kinetic-balance relation. Besides providing a natural connection between non-relativistic and relativistic orbitals, it guarantees that the correct non-relativistic kinetic-energy operator is recovered when the relativistic problem is represented in a finite basis.

The situation appears different in the Weyl representation. The unitary transformation connecting the conventions adopted here for the Dirac and Weyl representations can be written as\cite{Karwowski101002qua21053}
\begin{equation}
\label{eq:dirac_weyl_transform}
U_{D\rightarrow W} = \frac{1}{\sqrt{2}}
\begin{pmatrix}
\mathbb{I}_2 & -\mathbb{I}_2\\
\mathbb{I}_2 & \phantom{-}\mathbb{I}_2
\end{pmatrix},
\end{equation}
such that
\begin{equation}
\Psi_W = U_{D\rightarrow W}\Psi_D =
\frac{1}{\sqrt{2}}
\begin{pmatrix}
\psi_L^D-\psi_S^D\\
\psi_L^D+\psi_S^D
\end{pmatrix}
\equiv
\begin{pmatrix}
\psi_L\\
\psi_R
\end{pmatrix}.
\end{equation}
Thus, unlike in the Dirac representation, neither chiral component vanishes in the non-relativistic limit. Instead,
\begin{equation}
\label{eq:weyl_nr_guess}
\Psi_W^{\mathrm{NR}}
=
\frac{1}{\sqrt{2}}
\begin{pmatrix}
\Phi\\
\Phi
\end{pmatrix}.
\end{equation}
For a spin-up scalar orbital $\varphi$, this corresponds explicitly to
\begin{equation}
\Psi_W^{\mathrm{NR}}
=
\frac{1}{\sqrt{2}}
(\varphi,0,\varphi,0)^T.
\end{equation}
Hence, in the non-relativistic regime, the electronic state is distributed equally between the two chiral sectors rather than being localized in a single large component.

Although Eq.~\eqref{eq:weyl_nr_guess} represents the correct zeroth-order limit, it is not sufficient as a finite-$c$ approximation to a positive-energy Dirac spinor. This can be seen directly from the Weyl Hamiltonian. For equal Left and Right components, the expectation values of the two diagonal kinetic terms $c\hat{\pi}$ and $-c\hat{\pi}$ cancel. The non-relativistic kinetic energy therefore cannot be recovered from the zeroth-order spinor alone. Instead, it arises from the first-order difference between the two chiral components, in the same way that it arises through the coupling between the large and small components in the Dirac representation.

Using the convention introduced above,
\begin{equation}
\hat{\pi}= -\boldsymbol{\sigma}\cdot\hat{\mathbf p},
\end{equation}
Eq.\eqref{eq:dirac_kinetic_balance} becomes
\begin{equation}
\psi_S^D = -\frac{\hat{\pi}}{2mc}\Phi +\mathcal{O}(c^{-2}).
\end{equation}
Applying Eq.\eqref{eq:dirac_weyl_transform} to the kinetically balanced Dirac spinor therefore gives
\begin{equation}
\label{eq:weyl_balanced_guess}
\Psi_W^{\mathrm{KB}} = \frac{1}{\sqrt{2}}
\begin{pmatrix}
\left(1+\dfrac{\hat{\pi}}{2mc}\right)\Phi\\
\left(1-\dfrac{\hat{\pi}}{2mc}\right)\Phi
\end{pmatrix}
+\mathcal{O}(c^{-2}),
\end{equation}
or, equivalently,
\begin{equation}\label{guess}
\psi_{L/R}^{(0)} = \frac{1}{\sqrt{2}} \qty( 1\pm\frac{\hat{\pi}}{2mc})\Phi.
\end{equation}

This expression provides the natural Weyl analogue of kinetic balance. The two chiral components remain equal at leading order, while their first-order difference contains the momentum dependence required to reproduce the non-relativistic kinetic energy. Importantly, Eq.~\eqref{guess} should not be interpreted as an additional approximation specific to the Weyl formulation: it is simply the conventional kinetically balanced positive-energy Dirac spinor expressed in the Weyl representation.

This observation provides a convenient prescription for constructing the initial guess employed throughout this work. Non-relativistic orbitals are first obtained from the corresponding Schrödinger or Hartree–Fock problem. In a conventional four-component calculation they are embedded through Eq.\eqref{eq:dirac_balanced_guess}, whereas in the Weyl formulation the same orbitals are transformed according to Eq.\eqref{guess}. The two calculations therefore originate from the same physical non-relativistic states, differing only in their spinorial representation.

\section{Algorithm Development}

\subsection{Multiwavelets in Quantum Chemistry}

Although the overwhelming majority of development and applications in quantum chemistry is using \ac{AO} basis sets, multiwavelets have shown remarkable results, due to a combination of sound mathematical properties, efficient numerical algorithms and semantic simplicity.

The construction of multiwavelets, in their practical realization of Alpert\cite{Alpert1011370524016}, begins with a set of orthonormal polynomials defined $[0,1]$ (e.g. Legendre polynomials up to a given degree $k$). The basis can then be dyadically refined as follows:

\begin{equation}
    \scaling_{il}^n(x) = 2^{n/2} \scaling_i(2^n x - l)
\end{equation}

where $\scaling_{il}^n(x)$ is the $i$-th polynomial $(i=0 \ldots k)$ supported on the interval $[l/2^n, (l+1)/2^n]$ at scale $n = 0, 1, 2 \ldots..$

This constructions generates a nested sequence of scaling vector spaces $V_0 \subset V_1  \subset V_2 \ldots$ which are dense in the $L^2$ sense. The corresponding wavelet functions $\wavelet_{il}^n(x)$ are obtained as the orthogonal complement of two successive scaling spaces. A generic function can be represented either by making use of the scaling basis at the finest scale $N$ (reconstructed representation) or all the scaling functions at the coarsest scale and the tree of wavelet functions up to the next finest scale $N-1$

\begin{equation}
    f(x) \simeq \sum_{i=0}^k \sum_{l=0}^{2^N-1} s_{il}^n \scaling_{il}^n(x) = \sum_{i=0}^k s_{i}^0 \scaling_{i}^0(x) + \sum_{i=0}^k \sum_{n=0}^{N-1} \sum_{l=0}^{2^n-1} d_{il}^n \wavelet_{il}^n(x)
\end{equation}

where the coefficients $s_{il}^n$ and $d_{il}^n$ are obtained by integration via quadrature.

Rather than providing a comprehensive overview, we will here highlight the main fundamental properties of a \ac{MW} basis which makes the approach attractive and we refer to the literature for details on them.

\begin{description}
    \item[Completeness]: A MW basis can achieve completeness in the $L^2$ sense in two complementary ways: increasing the refinement or increasing the polynomial order. This provides a tunable and effective way to represent the required functions (orbitals, densities, potentials) with reduced complexity.\cite{Alpert1011370524016}
    \item[Disjoint support and adaptive refinement]: Functions (orbitals, densities, potentials) are represented on a three-dimensional grid covering the whole discretized space. Functions on different nodes have disjoint support (zero overlap), which enables adaptive refinement only where necessary to achieve the precision requested.\cite{Alpert101006jcph20027160}
    \item[Vanishing moments and compact representation]: A multiwavelet basis is constructed as the orthogonal complement of the so-called multiscaling basis in between two successive scale. As such, it is by construction orthogonal to polynomials up to a given order within its own support. Therefore the representation in a \ac{MW} basis of a smooth function is rapidly converging. Combined with adaptive refinement (point above) this allows for sparse representations of functions.\cite{Alpert1011370524016}
    \item[Non-Standard form of operators]: The so-called \ac{NS} form of operators is employed. The main advantage is represented by an effective decoupling of scales when operators are applied. This provides an automatic separation of long- and short-range effects for all required operators, without ad-hoc, operator specific techniques such as the Fast Multiple Method.\cite{Beylkin101016jacha200708001, 10108000268970412331319236, Frediani2013_Helmholtz_poissan}
    \item[Separated representation of operators]: Operator kernels such as the Poisson ($1/r$) or the Helmholtz kernel ($\exp{-\mu r}/r$) are decomposed in a sum of Gaussian functions. Their separability is exploited to reduce the formal scaling of operator applications.\cite{Beylkin101016jacha200701001}
    \item[Error control and precision-driven calculations]: The \ac{MW} framework enables rigorous error estimates for all the operations required (function projection, operator application, linear algebra, function multiplication, etc...). Therefore all operations can be carried out up to the required precision and no further, limiting the computational overhead to the minimum required.\cite{10108000268970412331319236, Frediani2013_Helmholtz_poissan, Jensen101021acsjpclett7b00255}
    \item[Narrow semantic gap]: In contrast to standard \ac{AO} methods, the sematic gap between the governing equations (e.g. the \ac{KS} equations of \ac{DFT}) and the code implementation is limited. This allows faster method development and exploration of avenues which are otherwise precluded by the limitation of the \ac{AO} implementation\cite{bjorgve_vampyr_2024, Harrison10113715m1026171}. This last point is central to this work as will be shown in the rest of the exposition.
\end{description}

\subsection{One-Electron Case}
To illustrate how the Weyl representation leads to a simplification in the working equations, especially within a \ac{MW} framework, we first derive the equations for the simpler one-electron case.

Applying the Weyl Hamiltonian \eqref{eq.Weyl_Hamiltonisn} to the chiral spinor components yields the coupled system of equations:
\begin{equation}\label{eq:Weyl_System}
    \begin{cases}    
      E \; \psi_L(x) &= \hat{V}\; \psi_L(x) + c \hat{\pi}\; \psi_L(x) + mc^2\; \psi_R(x)\\
      E \; \psi_R(x) &= \hat{V}\; \psi_R(x) - c \hat{\pi}\; \psi_R(x) + mc^2\; \psi_L(x)
    \end{cases},
\end{equation}
where each row represents a two-component complex spinor equation and we have introduced the  helicity-scaled momentum operator $\hat \pi = -(\vec \sigma\cdot \hat p)$.

The first equation contains $\psi_R(x)$ with a simple multiplicative factor and can hence be employed to obtain the energy-dependent coupling operator $\hat{R}(E)$:
\begin{equation}\label{eq:R-Weyl}
    \psi_R = \hat{R}^{W}(E)\psi_L, \quad \text{where} \quad \hat{R}^{W}(E) = -\frac{1}{mc^2}\qty((\hat{V}-E)+ c \hat{\pi}).
\end{equation}
It is worth noticing that in the Dirac representation the term $V-E$ is at the denominator of $\hat{R}^{D}(E)$, whereas for the Weyl representation the corresponding term is at the numerator. This is a significant advantage as we shall demonstrate. 

By making use of Eq.~\ref{eq:R-Weyl} into the second equation of \eqref{eq:Weyl_System} we obtain:
\begin{equation*}
    \psi_L = \frac{1}{m^2c^4} \qty{ (\hat{V}-E)^2 - c \qty[\hat{\pi}, (\hat{V}-E)] - c^2 \hat{\pi}^2}\psi_L.
\end{equation*}
Utilizing the Pauli matrix identity $\hat{\pi}^2 = (\vec{\sigma}\cdot\vec{p})(\vec{\sigma}\cdot\vec{p}) = p^2 = -\nabla^2$, and noting that the scalar energy $E$ commutes with $\hat{\pi}$, we obtain:
\begin{equation}\label{propagator}
    \psi_L(x) = \frac{1}{m^2c^4} \qty{\hat{V}^2 - 2E \hat{V} + E^2 - c \qty[\hat{\pi},\hat{V}] +c^2 \nabla^2} \,\psi_L(x).
\end{equation}
The latter equation can be rearranged to obtain a Helmholtz-type differential operator on the left-hand side:
\begin{equation}\label{eq:2_component_diff_eq}
    \qty[-\nabla^2 + \qty(\frac{m^2c^4-E^2}{c^2})]\psi_L(x) = \frac{1}{c^2} \qty{\hat{V}^2 - 2E \hat{V} - c \qty[\hat{\pi},\hat{V}] }\psi_L(x).
\end{equation}
This equation is equivalent to the propagator used by Karwowski \textit{et al.} \cite{Karwowski101002qua21053},
which matches the standard inhomogeneous Helmholtz equation $(-\nabla^2 +\mu^2 )f(x) = g(x)$, where the screening parameter is defined as $\mu^2 = \frac{m^2c^4-E^2}{c^2}$. The formal inverse of the operator can be applied via convolution with the Helmholtz Green's function kernel $\hat{G}^\mu$:
\begin{equation}
    \hat{G}^\mu \star f(r) = \int \frac{e^{-\mu \abs{r-r'}}}{4\pi \abs{r- r'}} f(r) \ \dd r = [-\nabla^2+\mu^2]^{-1}f(r).
\end{equation}

\paragraph{External Potential Simplification}
For a static external potential $\hat{V}$ (where $\hat{p}(\hat{V}\psi) = \hat{p}(\hat{V})\psi + \hat{V}\hat{p}\psi$), the commutator term simplifies to:
\begin{equation}\label{eq:V_simplification}
    -c \qty [(\vec{\sigma}\cdot \vec{p}), \hat{V}] {\psi} = ic\  (\vec \sigma\cdot \Vec{\nabla} \hat V){\psi}.
\end{equation}

\paragraph{Final Iterative Scheme}
Incorporating this simplification yields the final equation for the iterative propagation of the Left component:
\begin{equation}\label{eq:1el_prop}
    \psi_L ^{n+1} = \hat{G}^{\mu_n}\star \frac{\hat{\mathcal{V}}}{c^2}\psi_L^n,
\end{equation}
where
\begin{equation}\label{eq:potential-one-el}
\hat{\mathcal{V}}= \hat{V}^2- 2 E\hat{V } + i c \; (\vec{\sigma}\cdot \Vec{\nabla}\hat{V}). 
\end{equation}
The final equation has the same structure as the integral form of the one-electron Schrödinger equation which is commonly used in a \ac{MW} framework. This is however a 2-component equation: the coupling between the two components is given the last term. Moreover, in contrast to the non-relativistic case, the right hand side of Eq.~\ref{eq:1el_prop} is now energy-dependent.
A closer connection can be drawn if one separates the total energy as $E = mc^2 + \epsilon$. The final equation can then be rewritten as:
\begin{equation}
    \left( 
    -\nabla^2 - 2m\epsilon - \frac{\epsilon^2}{c^2} 
    \right) \orbleft = -2 V \left(
    m - \frac{V}{2c^2} + \frac{\epsilon}{c^2} + \frac{i}{2cV} \sigma \cdot \nabla V
    \right)\orbleft
\end{equation}
All relativistic corrections vanish in the $c\rightarrow +\infty$ limit and the usual non-relativistic equation for the electron $(-\nabla^2 - 2 \epsilon)\psi = -2V\psi$, is recovered by setting $m=1$.

\subsection{N-Electron Case}
In analogy with the non-relativistic case~\cite{10108000268970412331319236, Jensen2022_integral_formalism}, the formalism can be extended to the many-electron framework. In this work we limit ourselves to the Dirac--Fock equation, and the instantaneous Coulomb interaction $1/r$. Extensions to \ac{DFT}, correlated methods or retardation effects such as the Breit Hamiltonian will be considered in future work. 
In order to simplify the notation we will implicitly assume the 4-component or 2-component nature of the equations in all the following derivation, relying on context to discriminate between 4- and 2-component cases.

\begin{equation}
    [\hat{\mathcal{F}}\Psi]_i = \sum_{j=1}^N F_{ji} \Psi_j,
\end{equation}
In the above Equation $\hat{\mathcal{F}}=\hat{h}_D + \hat{V} + \hat{J} - \hat{K}$ is the 4-component Fock operator which contains the Dirac Hamiltonian $\hat{h}_D$, the external potential $\hat{V}$, the Hartree potential $\hat{J}$, and the exchange interaction $\hat{K}$.



By making use of the Weyl representation the equation can be written as
\begin{equation}\label{eq:Kohn-Sham_matrix}
    \hat{\mathcal{F}}\Psi_i = F_{ii} \Psi_i + \sum_{j\not=i}F_{ji}\,\Psi_j = \begin{pmatrix}
        c \hat{\pi} + \hat{V}  & mc^2 \\ mc^2 & - c\hat{\pi}+\hat{V} 
    \end{pmatrix}
    \begin{pmatrix}
        \orbleft_i \\ \orbright_i 
    \end{pmatrix}.
\end{equation}
Isolating the $R$ component yields the many-electron coupling operator:
\begin{equation}\label{eq:exact_coupling}
    \orbright_i = -\frac{1}{mc^2}
    \qty{
    \qty[
    (\hat{V}-F_{ii})+ c \hat{\pi}
    ] 
    \orbleft_i - \sum_{j\not=i}F_{ji} \orbleft_j}.
\end{equation}
The coupling operator in the many-electron case is now dependent on all occupied spinors, unless a canonical representation (diagonal Fock matrix) is enforced. By making use of Eq.~\ref{eq:Kohn-Sham_matrix} \& ~\ref{eq:exact_coupling}
\begin{equation}\label{eq:exact_N_el_Prop}
    \orbleft_i = \frac{1}{c^2}\hat{G}_{\mu_i}\star 
    \left[{\hat{\mathcal{V}}}\orbleft_i 
    +2\sum_{j\not=i } F_{ji}\left( \frac{F_{ii}+F_{jj}}{2}-\hat{V} \right)\orbleft_j +\sum_{j\not=i}\sum_{k\not=j} F_{ji}F_{kj} \orbleft_k\right],
\end{equation}
where the effective potential operator must retain the full commutator format:
\begin{equation}\label{eq:potential-many-el}
    \hat{\mathcal{V}_i}= (\hat{V})^2- 2 F_{ii}\hat{V } -  c \; \left[\hat{\pi}, \hat{V}\right].
\end{equation}

\paragraph{Effective Potential Simplification Scheme}
Evaluating the commutator $\qty[\hat{\pi}, \hat{V}]$ when involving the non-local exchange operator $\hat{K}$ is computationally demanding. We can bypass this bottleneck by rewriting the action of $\hat{\mathcal{V}}$ on $\orbleft_i$ and use Eq.~\ref{eq:V_simplification}:
\begin{equation}
    \hat{\mathcal{V}}\orbleft_i = \hat{V}\qty(\hat{V} -F_{ii}+c\hat{\pi})\orbleft_i - F_{ii} \hat{V}\orbleft_i -  c \hat{\pi} \hat{V}\orbleft_i.
\end{equation}
The term in parenthesis matches the operator from the exact coupling expression \eqref{eq:exact_coupling}, and we can substitute it to incorporate the opposite-chirality component. This leads to a more efficient formulation, which only requires the evaluation of the direct potential actions $\hat{V}\orbleft_i$ and $\hat{V}\orbright_i$:
\begin{equation}\label{eq:simplified_pot}
    \hat{\mathcal{V}}\orbleft_i = -mc^2 \,\hat{V}\orbright_i  - F_{ii} \, \hat{V}\orbleft_i -  c \hat{\pi} \, \hat{V}\orbleft_i  + \sum_{j\not=i} F_{ji}\,\hat{V}\orbleft_i.
\end{equation}

\paragraph{Iteration and convergence}
As for the non-relativistic case, the integral equation~\ref{eq:exact_N_el_Prop}, can be iterated until convergence. The additional complication resides in the coupling between the left and right components. The right component is needed to compute the Fock matrix elements $F_{ji}$. 
Although the need to compute $\orbright_i \ (i=1,n)$ could in principle be bypassed by exploiting the coupling relationship, in this first implementation we have for simplicity chosen to compute it explicitly, in a two-step procedure.

\subsection{Variational collapse}

In the one-electron case, given $V$ to be the homogeneous, spherically symmetric Coulomb potential, one can simplify the differential operator shown in Eq.~\eqref{eq:2_component_diff_eq}, removing the energy dependence and turning it into a proper Hamiltonian. To do this, we rescale the coordinates of the wavefunctions according to the following mapping \cite{Karwowski101002qua21053}
\begin{equation}\label{eq:r_mapping}
    \vec{r}\mapsto
    \qty(1+\frac{E}{mc^2})\vec{r}
    \equiv \vec{\rho}.
\end{equation}
This transformation results in the following Hamiltonian:
\begin{equation}\label{eq:2c-hamiltonian_w}
    H_w^\pm =
    \frac{\hat{\pi}^2}{2m}
    + \hat V(\vec{\rho})
    - \frac{\hat V(\vec{\rho})^2}{2mc^2}
    \mp \frac{1}{2mc}\comm{\hat{\pi}}{V(\rho)}.
\end{equation}
As shown in Ref.~\cite{Karwowski101002qua21053}, this satisfies
\begin{equation}
    H_w^\pm(\vec{\rho})\psi_{L/R}(\vec{\rho})
    =
    \epsilon\,\psi_{L/R}(\vec{\rho}),
\end{equation}
where the eigenvalues of this equation are related to the energy of the original Dirac equation by
\begin{equation}
    \epsilon =
    E\qty(1+\frac{E}{2mc^2})
    \qty(1+\frac{E}{mc^2})^{-2}.
\end{equation}
Thus, each eigenvalue of $H_w^\pm$ can be associated with the corresponding Dirac energy through the mapping $\epsilon_i\leftrightarrow E_i$.

The important consequence of this transformation is that the two-component Hamiltonian $H_w^\pm$ has a spectrum bounded from below. In particular, both the positive- and negative-energy continua of the Dirac Hamiltonian are mapped onto the positive spectrum of $H_w^\pm$, while the bound states are mapped onto its discrete negative spectrum \cite{Karwowski101002qua21053}. The physically relevant bound states therefore belong to a part of the transformed spectrum that is bounded from below. In this one-electron Coulomb case, the Weyl representation combined with the energy-dependent coordinate scaling thus maps the unbounded Dirac spectral problem onto a Schrödinger-like problem for which the usual variational minimization can be applied without variational collapse.

This argument is, however, as already stated before is restricted to the one-electron Coulomb problem and cannot be directly extended to the $N$-electron case. The transformation in Eq.~\eqref{eq:r_mapping} relies on the homogeneity of the Coulomb potential, whereas the effective many-electron operators $\hat J$ and $\hat K$ are not homogeneous as they stem from multiple sources, with $\hat K$ being additionally nonlocal. Consequently, one cannot simply substitute
\begin{equation}
    \hat V = \hat V_{ne}+\hat J-\hat K
\end{equation}
into Eq.~\eqref{eq:2c-hamiltonian_w} to obtain an equivalent Fock-like Hamiltonian with the same spectral properties. Nonetheless, the here presented SWORD (Spinorial Weyl Operator Reduction of Dirac) method offers a different way of distinguishing the positive- and negative-energy branches during the iterative procedure.

When performing calculations with a four-component Hamiltonian, variational collapse originates from the fact that the Dirac spectrum is not bounded from below. A direct minimization of the Rayleigh quotient can therefore lower the energy by mixing the target electronic state with states belonging to the negative-energy continuum, unless additional constraints or projection procedures are introduced. In SWORD, instead, only one Weyl component is explicitly propagated, while the complementary component is reconstructed through the energy-dependent balance relation in Eq.~\eqref{eq:exact_coupling}. The energy branch can therefore be explicitly encoded in the reconstruction step rather than allowing an unconstrained variation over the complete four-component space.

For the $L$ component, the dependence on $E$ enters through a linear term in the propagator of Eq.~\eqref{eq:exact_N_el_Prop}. A more explicit distinction between the two energy branches appears in the balance equation, Eq.~\eqref{eq:exact_coupling}, through the term proportional to $E/(mc^2)$. For a positive-energy electronic state, the energy can be written as
\begin{equation}
    E=mc^2+\varepsilon,
\end{equation}
where $\varepsilon$ is the energy relative to the positive-energy rest-mass threshold. Consequently,
\begin{equation}
    \frac{E}{mc^2}\psi_L
    =
    \qty(1+\frac{\varepsilon}{mc^2})\psi_L.
\end{equation}
For a state belonging to the negative-energy branch, one instead writes
\begin{equation}
    E=-mc^2+\varepsilon,
\end{equation}
giving
\begin{equation}
    \frac{E}{mc^2}\psi_L
    =
    \qty(-1+\frac{\varepsilon}{mc^2})\psi_L.
\end{equation}
The two branches can therefore be explicitly distinguished through the sign of their rest-mass contribution in the balance relation. By selecting the appropriate reference branch, $+mc^2$ for electronic states or $-mc^2$ for negative-energy states, the reconstruction step retains information about which part of the Dirac spectrum is being targeted. This branch dependence provides SWORD with a mechanism for restricting the iterative procedure to the desired sector of the spectrum, rather than relying on an unrestricted minimization of the four-component Dirac Hamiltonian.

\subsection{Comparison with other \ac{MW} implementations}
The first \ac{MW} implementation of the Dirac Hartree-Fock method has been presented by Anderson et.al.\cite{Anderson10106315128908}. The central piece of their work consisted in using a Green's function formulation for the level-shifted Dirac Hamiltonian, in close analogy with the non relativistic case. 
\begin{equation}\label{eq:dirac-scf-convolution}
    \Psi_{i} = - \frac{1}{c^{2}} \left[
    (\hat{h}_{D} + F_{ii}) 
    \hat G_{\mu_i}
    \right]
    \star
 \left[ \hat{V} \Psi_{i} -\sum_{j\neq i} F_{ji}\Psi_j \right].
\end{equation}
However such a Green's function is constituted by both an integral kernel which is well suited for a \ac{MW} treatment, and a differential operator, which is only defined in a weak sense and can generate numerical noise\cite{Anderson101016jjcpx2019100033}.

As originally shown by Kutzelnigg and Wallheimer~\cite{VAR_Collapse}, and implemented recently in our group in a \ac{MW} framework~\cite{masotti2026}, one can avoid the differential operator, by making use of the squared Dirac Hamiltonian:
\begin{equation}\label{eq:dirac-squared-scf-convolution}
    \Psi_i 
    =
    -2m
    \hat G_{i}^{\mu_i}
    \star
    \left[
    \hat{\mathcal{V}}\Psi_i
    +
    \frac{1}{2mc^2}\sum_{j\neq i}(\hat{\mathfrak{D}}^{2})_{ij}\Psi_j
    \right],
\end{equation}
where the matrix elements of $\hat{\mathfrak{D}}^{2}$ operator are obtained as:
\begin{equation}\label{eq:matrix-product}
(\hat{\mathfrak{D}}^{2})_{ij} \simeq \sum_k D_{ik}D_{kj}.
\end{equation}
This strategy leads to a 4-component equation which is structurally equivalent to the non-relativistic Schrödinger equation, and can be propagated using the bound state Helmholtz kernel exactly as for the non-relativistic case.

However, the term
\begin{equation}\label{eq:generalized-potential}
   \hat{\mathcal{V}} = 
    \frac{[\beta, \hat{V}]_{+}}{2}
    + \frac{[\Vec{\alpha}\cdot \Vec{p}, \hat{V}]_{+}}{2mc}
    + \frac{\hat{V}^{2}}{2mc^2}I_{4},
\end{equation}
depends on the square of the potential, which requires once again to compute a substantial amount of convolutions to evaluate the action of the two-electron interaction.


The advantage of this method is to avoid the use of a differential operator after the application of the Helmholtz kernel, thus providing additional numerical precision.
It can also be shown that if one applies the decoupling operator $R^W(E)$ to the squared Dirac Hamiltonian equation, the same two-component equations presented in this work are obtained. 

The present method provides the additional benefit of a propagator equation which both avoids the differentiation step, and also requires only a two-component equation. 
The computational advantage is obvious: the time-consuming propagation step is reduced by a factor two.
In conclusion, the 2-component method discussed in this article can offer an important speed up in performance for the treatment of relativistic effects, while at the same time avoiding any approximation, because the method is completely equivalent to a 4-component treatment. 

\subsection{Code structure}
The algorithms presented above have been implemented in a development version of the \remrchem{} code, which is a Python software based on the \vampyr{} library which in turns exposes the \ac{MW} framework implemented in \mrcpp{}. Our starting point was the code developed for \citenum{masotti2026}. The SCF iteration was developed starting from the 4-component one used in \citenum{masotti2026}, developing the required infrastructure for the chiral two-component method described in the present work. 

Using \vampyr{} as the main development environment, allows for great flexibility but on the other hand, it limits the size of systems whcih can be treated. That is why we will limit the results presented in the following section to 1- 2- and 4-electron systems. The aim is to provide a proof-of-concept implementation, and to demonstrate the correctness and feasibility of the approach. Additionally, to simplify the calculations, we exploited Kramers' time reversal symmetry \cite{KTRS}.

\paragraph{The main algorithmic blocks}

The workflow of the implementation, irrespective of the number of electrons, proceeds through the following sequence of mail logica blocks:
\begin{enumerate}
    \item \textbf{Initialization:} The initial set of 4-component spinors is constructed by making use of non-relativistic orbitals. They are partitioned into Left and Right components by making use of the kinetic balance, and orthonormalizes through a Löwdin symmetric orthogonalization procedure.
    \item \textbf{SCF warm-up:} Before the first with the 2-component method, it is necessary to smoothen out high frequency numerical noise coming from the employment of derivatives on the non-relativistic guess function with a suitable low-pass filter.
    \item \textbf{Mean-Field Operators:} At each iteration, the set of spinors is used to compute the current Coulomb ($\hat{J}$) and Exchange ($\hat{K}$) potentials.
    \item \textbf{Fock Matrix evalation:} The Fock matrix elements $F_{ji} = \mel{\Psi_i}{\hat{\mathcal{F}}}{\Psi_j}$ are computed on the 4-component spinors.
    \item \textbf{Left Component Propagation:} The Left components $[\orbleft_i]^{(n)}$ are propagated as shown in Eq.~\ref{eq:exact_N_el_Prop}.
    \item \textbf{Right Component Balancing:} The Right components $[\orbright_i]^{(n+1/2)}$ are obtained from the propagated Left components and the previous iterate of the Fock matrix $F^{{n}}$, through the exact coupling relation \eqref{eq:exact_coupling}. 
    \item \textbf{Reconstitution and Orthogonalization:} The full 4-component spinors $\Psi_i^{n+1}$, are orthonormalized, and convergence cis evaluated against a predefined threshold.
\end{enumerate}

This algorithm provides an exact formulation where the sole source of numerical approximation lies within the controlled multiwavelet convolution. The intermediate notation $\psi_R^{n+1/2}$ indicates that the coupling utilizes the energy from step $n$.  

In general, using the previous step's energy to find $\psi_R^{n+1/2}$ from $\psi_L^{n+1}$ does not introduce any approximations. Taking Eq. \eqref{eq:exact_coupling} we can show that 
\begin{equation}    
\norm{\psi_R^{n+1/2}-\psi_R^{n+1}}\propto \frac{\epsilon^{n}-\epsilon^{n+1}}{mc^2}
\end{equation}
Thus in the limit of convergence $\psi_R^{n+1/2}\equiv \psi_R^{n+1}$ within the numerical precision.

As opposed to other common 2-component methods, SWORD's coupling operator as well as the propagator explicitly depend on the energy for any given orbital. Thus each different electron in the system is propagated iteratively with a slightly different equation, in a similar manner to the Cowan-Griffin method \cite{Cowan:76}.

To better illustrate the SCF cycle, figure \ref{fig:Algorithm_visual} shows a typical computation starting form the non-relativistic starting guess.
\begin{figure}[H]
    \centering
\resizebox{0.8\textwidth}{!}{%
\begin{circuitikz}
\tikzstyle{every node}=[font=\LARGE]
\draw  (6.25,20) rectangle  node {\LARGE $\ket{\psi_L^n}$,  $E^n$} (10,15);
\draw  (12.5,17.5) -- (14.375,16.75) -- (16.25,17.5) -- (14.375,18.25) -- cycle;
\node at (14.375, 17.5) {$\hat{G}(E^n)$};
\draw  (20.5,17.5) ellipse (1.75cm and 1cm);
\draw  (28.75,17.5) ellipse (1.75cm and 1cm);
\node at (20.5,17.5) {$\ket{\psi_L^{n+1}}$};
\node at (28.75,17.5) {$\ket{\psi_R^{n+\frac{1}{2}}}$};
\draw [->, >=Stealth] (10,17.5) -- (12.5,17.5);
\draw [->, >=Stealth] (16.25,17.5) -- (18.75,17.5);
\draw [->, >=Stealth] (22.25,17.5) -- (27,17.5)  node[midway, above] {$\hat{R}(E^n)$};
\draw [short] (20.5,16.5) -- (20.5,14.5);
\draw [short] (28.75,16.5) -- (28.75,14.5);
\draw  (23.5,15.25) rectangle  node {$E^{n+1}$} (25.75,13.5);
\draw [->, >=Stealth] (20.5,14.5) -- (23.5,14.5);
\draw [->, >=Stealth] (28.75,14.5) -- (25.75,14.5);
\draw [->, >=Stealth, dashed] (17.5,15.75) -- (10,15.75) node[midway, below] {Next iteration};
\draw [dashed] (17.5,15.75) -- (19.5,13.75);
\draw [->, >=Stealth, dashed] (19,17) -- (17.5,15.75);
\draw [->, >=Stealth, dashed] (23.5,13.75) -- (19.5,13.75);
\end{circuitikz}
}%
    \caption{General representation of a 1 electron algorithm.}
    \label{fig:1el_alg}
\end{figure}
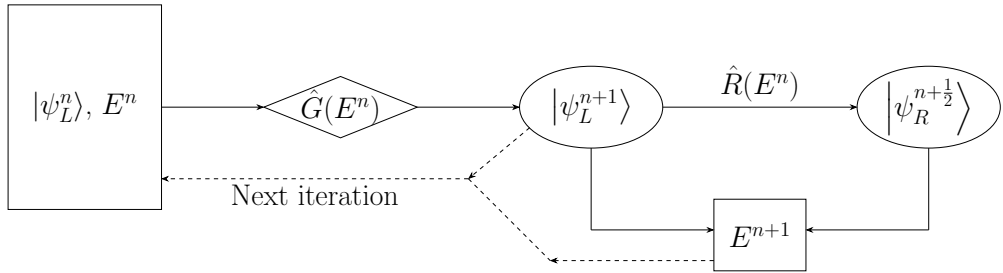
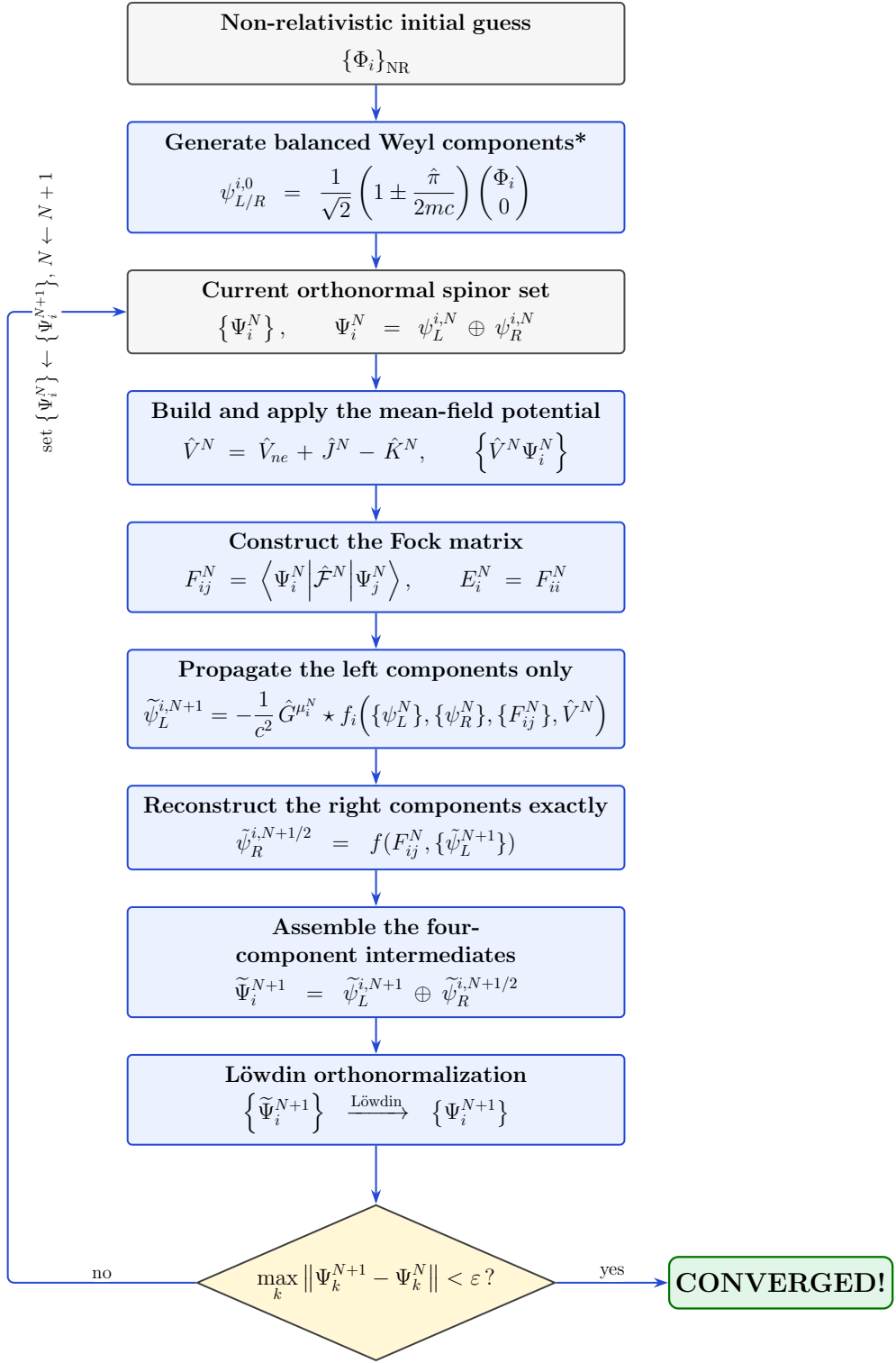
\begin{figure}[h!]
    \centering
    \resizebox{0.8\textwidth}{!}{
    \begin{tikzpicture}[node distance=8mm and 22mm]

\node[state] (nr) {
  \textbf{Non-relativistic initial guess}\\[+2mm]
  $\left\{\Phi_i\right\}_{\mathrm{NR}}$
};

\node[process, below=of nr] (balance) {
  \textbf{Generate balanced Weyl components*}\\[+2mm]
  $\displaystyle
  \psi_{L/R}^{i,0}=\frac{1}{\sqrt{2}}
  \left(1\pm\frac{\hat \pi}{2mc}\right)
  \begin{pmatrix}\Phi_i\\0\end{pmatrix}$
};

\node[state, below=of balance] (spinorset) {
  \textbf{Current orthonormal spinor set}\\[+2mm]
  $\displaystyle
  \left\{\Psi_i^{N}\right\},\qquad
  \Psi_i^N=\psi_{L}^{i,N}\oplus\psi_{R}^{i,N}$
};

\node[process, below=of spinorset] (potential) {
  \textbf{Build and apply the mean-field potential}\\[+2mm]
  $\displaystyle
  \hat V^N=\hat V_{ne}+\hat J^N-\hat K^N,
  \qquad
  \left\{\hat V^N\Psi_i^N\right\}$
};

\node[process, below=of potential] (fock) {
  \textbf{Construct the Fock matrix}\\[+2mm]
  $\displaystyle
  F_{ij}^N=\left\langle\Psi_i^N\middle|\hat{\mathcal F}^N\middle|\Psi_j^N\right\rangle,
  \qquad E_i^N=F_{ii}^N$
};

\node[process, below=of fock] (propagate) {
  \textbf{Propagate the left components only}\\[+2mm]
  $\displaystyle
  \widetilde{\psi}_{L}^{i,N+1}
  =-\frac{1}{c^2}\,\hat G^{\mu_i^N}\star
   f_i\!\left(\{\psi_L^N\},\{\psi_R^N\},\{F_{ij}^N\},\hat V^N\right)$
};

\node[process, below=of propagate] (reconstruct) {
  \textbf{Reconstruct the right components exactly}\\[+2mm]
  $\displaystyle
  \tilde{\psi}_{R}^{i,N+1/2}
  = f(F_{ij}^N,\{\tilde \psi_L^{N+1}\})$
};

\node[process, below=of reconstruct] (assemble) {
  \textbf{Assemble the four-component intermediates}\\[+2mm]
  $\displaystyle
  \widetilde{\Psi}_i^{N+1}
  =\widetilde{\psi}_{L}^{i,N+1}
  \oplus
  \widetilde{\psi}_{R}^{i,N+1/2}$
};

\node[process, below=of assemble] (lowdin) {
  \textbf{Löwdin orthonormalization}\\[+2mm]
  $\displaystyle
  \left\{\widetilde{\Psi}_i^{N+1}\right\}
  \xrightarrow{\text{Löwdin}}
  \left\{\Psi_i^{N+1}\right\}$
};

\node[decision, below=13mm of lowdin] (test) {
  $\displaystyle
  \max_k\left\|\Psi_k^{N+1}-\Psi_k^N\right\|<\varepsilon\,?$ 
};

\node[terminal, right=25mm of test] (converged) {CONVERGED!};

\draw[flowarrow] (nr) -- (balance);
\draw[flowarrow] (balance) -- (spinorset);
\draw[flowarrow] (spinorset) -- (potential);
\draw[flowarrow] (potential) -- (fock);
\draw[flowarrow] (fock) -- (propagate);
\draw[flowarrow] (propagate) -- (reconstruct);
\draw[flowarrow] (reconstruct) -- (assemble);
\draw[flowarrow] (assemble) -- (lowdin);
\draw[flowarrow] (lowdin) -- (test);

\draw[flowarrow] (test.east) -- node[smalllabel, above] {yes} (converged.west);

\coordinate (loopturn) at ($(spinorset.west)+(-22mm,0)$);
\draw[flowarrow]
  (test.west)
  -- node[smalllabel, above] {no} ++(-42mm,0)
  |- node[smalllabel, pos=0.73, rotate=90, anchor=south] {
       set $\left\{\Psi_i^{N}\right\}\leftarrow\left\{\Psi_i^{N+1}\right\}$,
       $N\leftarrow N+1$
     }
  (spinorset.west);

\end{tikzpicture}
    }
    \caption{Schematic representation of the algorithm for the N-electron case. The asterisk in the generation of the balanced components refers to the use of the scf warm-up step to smoothen irregularities.}
    \label{fig:Algorithm_visual}
\end{figure}

\section{Results and discussion}

The numerical assessment was designed to test the proposed SWORD formulation by introducing the ingredients of the Dirac--Hartree--Fock problem progressively. One-electron calculations first isolate the Green's-function propagation and the reconstruction of the opposite chirality. Two-electron closed-shell systems then introduce a self-consistent Coulomb potential while retaining an effectively local mean field. Finally, four-electron closed-shell systems introduce the non-local Hartree--Fock exchange operator and therefore provide the first benchmark containing all of the ingredients considered in the present Dirac--Coulomb implementation.

All calculations were performed with the \vampyr{} Python library, using an adaptive multiwavelet (MW) representation of the electronic wavefunctions. The reference four-component implementation and SWORD were constructed to follow the same SCF workflow as closely as possible. In the former, all four complex spinor components are propagated by the Green's-function update, whereas in SWORD only the two components of one chirality are propagated and the opposite chirality is subsequently reconstructed through the coupling relation derived above. Both calculations were initialized from the same non-relativistic orbitals, embedded in the corresponding spinorial representation, and converged using identical criteria. This setup was chosen to isolate, as far as possible within the present proof-of-concept implementation, the numerical consequences of replacing four-component propagation by propagation and reconstruction in the Weyl representation.

For the one-electron systems, the initial orbital was the analytical non-relativistic hydrogenic solution projected onto the MW representation and embedded according to Eq.~\eqref{eq:weyl_balanced_guess}. For the many-electron systems, hydrogenic orbitals constructed with screened nuclear charges obtained from Slater's rules were used as the non-relativistic starting point and embedded in the same manner. Before the production SCF calculation, each starting guess was subjected to a short, low-precision four-component warm-up. The resulting smoothed orbitals were then used as common input to the Dirac and SWORD calculations. The numerical motivation for this preliminary step is discussed in Sec.~\ref{sec:warmup_considerations}.

All calculations used the speed of light $c=137.03599913900001$ in atomic units. Unless otherwise stated, the electron--electron interaction was restricted to the Dirac--Coulomb Hamiltonian; Breit and related non-instantaneous corrections were therefore not included. This choice keeps the comparison focused on the propagation strategy rather than on additional relativistic interaction terms.


\subsection{One-electron validation: local external potentials}

The one-electron problem provides the most direct validation of the reduced propagation scheme because it tests the Green's-function update and the coupling operator before any self-consistent electron--electron interaction is introduced. Five hydrogen-like systems spanning a broad range of nuclear charges were considered: \ce{H}, \ce{Ne^{9+}}, \ce{Kr^{35+}}, \ce{Xe^{53+}}, and \ce{Hg^{79+}}. For nuclei heavier than hydrogen, the nuclear charge was represented by a finite-size Fermi distribution parametrized according to Ref.~\cite{Visscher_1997}. For hydrogen, a point-charge model with numerical smoothing of the Coulomb singularity was employed\cite{Yanai2004}.

The calculations were repeated over the MW5--MW8 precision range. Validation was performed at two levels. First, SWORD was compared directly with the four-component MW implementation under matched numerical conditions. Second, the resulting energies were compared with independent \grasp{} reference calculations using the corresponding nuclear model. For the finite-nucleus calculations, this numerical reference is more appropriate than the analytical point-Coulomb spectrum because the latter describes a different nuclear model. Hydrogen additionally provides a point-Coulomb control case.

\begin{figure}[h]
    \centering
    \includegraphics[width=0.8\linewidth]{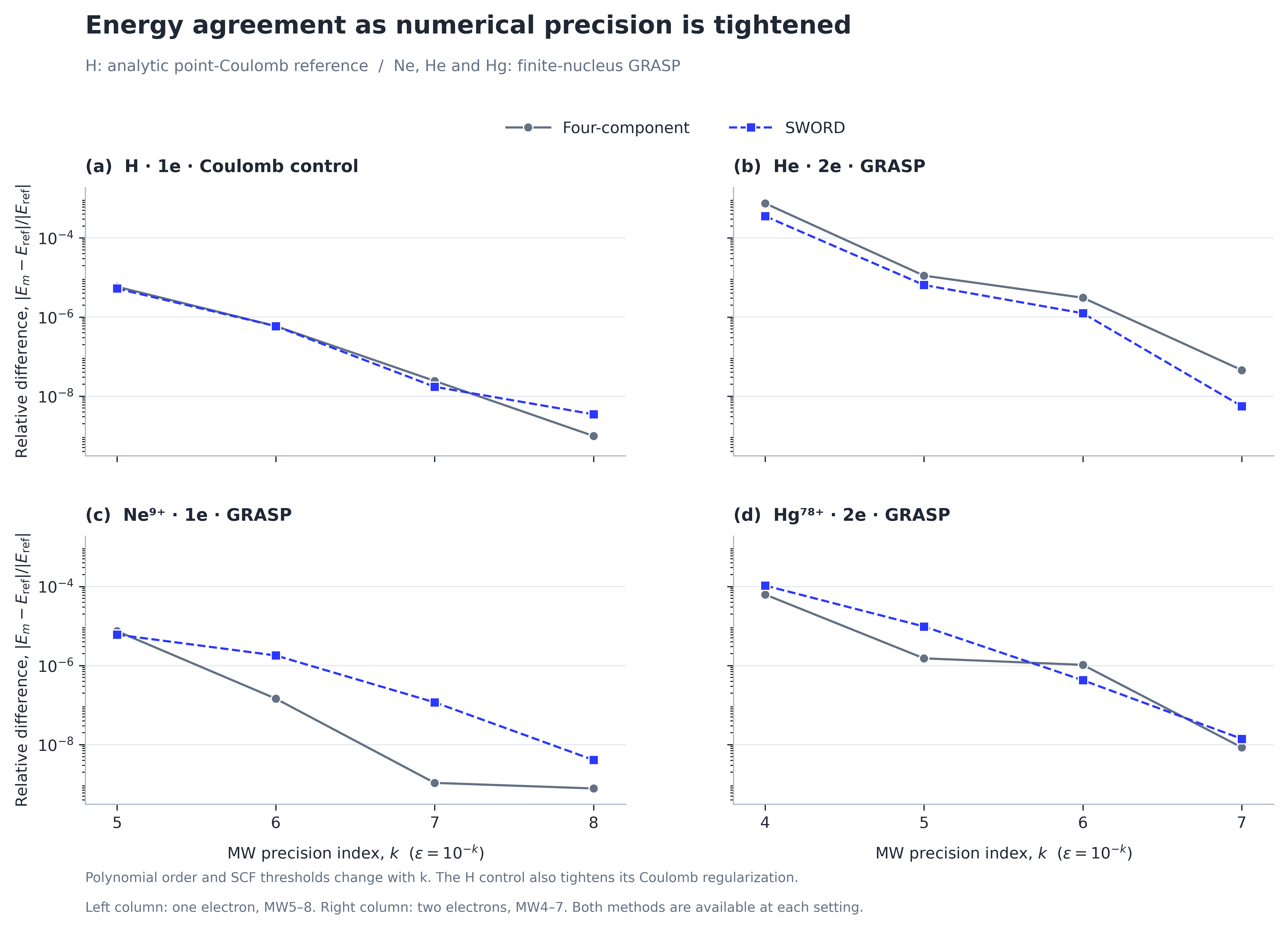}
    \caption{Relative total-energy error as the MW precision is tightened for representative one- and two-electron systems. The one-electron calculations span MW5--MW8 and the two-electron calculations MW4--MW7. Hydrogen is compared with the analytical point-Coulomb reference, whereas \ce{Ne^{9+}}, \ce{He}, and \ce{Hg^{78+}} are compared with finite-nucleus \grasp{} reference calculations. Four-component Dirac and SWORD results were obtained at matched numerical settings.}
    \label{fig:1el_2el_prec_sweeps}
\end{figure}

The precision dependence is summarized in Fig.~\ref{fig:1el_2el_prec_sweeps}, together with representative two-electron results discussed below. Across the tested precision range, both the four-component implementation and SWORD systematically approach the reference energy as the MW threshold is tightened. In particular, SWORD remains within the accuracy expected from the imposed numerical threshold, establishing that propagation of a single chirality followed by reconstruction of the second does not compromise the attainable energy accuracy in the one-electron case.

The one-electron calculations also expose the computational advantage of the reduced propagation most clearly. As shown in Fig.~\ref{fig:timing_1el_v_2el}, the production SCF stage is approximately a factor of two faster for the representative one-electron calculations, consistent with propagating only half as many spinor components. When setup and the common warm-up procedure are included, the reduction in total wall time is necessarily smaller, but remains substantial. The reconstruction of the opposite chirality contributes only a small fraction of the update cost, confirming that the principal computational saving originates from reducing the number of Green's-function propagations.

\subsection{Two-electron systems: self-consistent local potentials}

The two-electron closed-shell case introduces the next element of the Dirac--Hartree--Fock problem: a mean-field Coulomb potential that changes during the SCF procedure. Kramers time-reversal symmetry was exploited, as in Ref.~\cite{masotti2026}, so that only one member of the occupied Kramers pair had to be treated explicitly\cite{KTRS}. This provides a controlled test of SWORD in the presence of a self-consistent potential without yet introducing the non-local exchange coupling between distinct occupied spatial orbitals.

The same progression in nuclear charge was retained for the helium-like sequence, with \ce{He} replacing the hydrogenic starting point. Calculations were performed over the MW4--MW7 precision range and, as in the one-electron case, SWORD and four-component results were compared both with one another and with \grasp{} calculations obtained using matched physical models. The representative results in Fig.~\ref{fig:1el_2el_prec_sweeps} show that the two formulations retain comparable numerical accuracy as the requested MW precision is tightened.

\begin{figure}[h]
    \centering
    \includegraphics[width=0.8\linewidth]{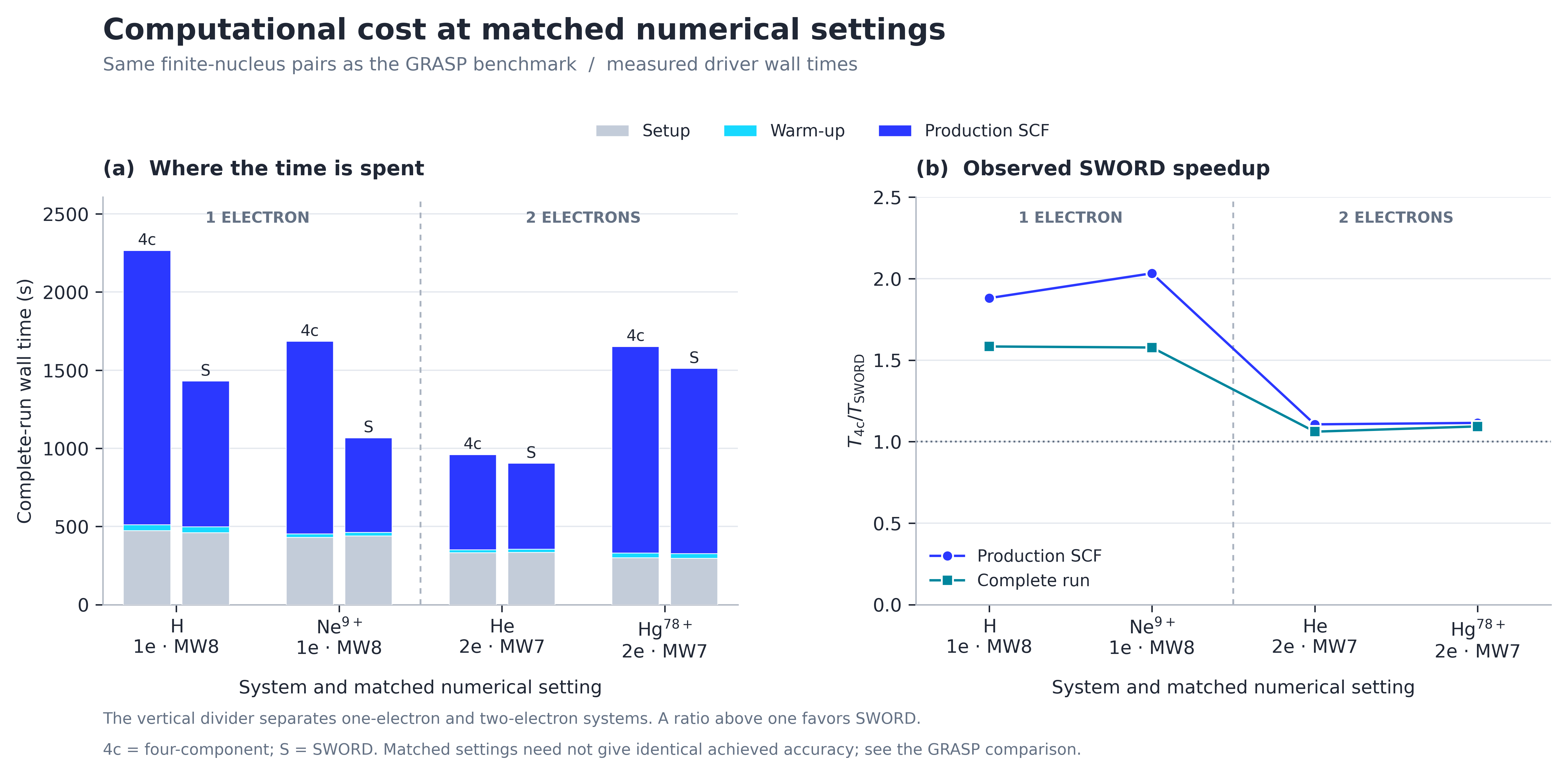}
    \caption{Computational cost of representative one- and two-electron calculations at matched numerical settings. Panel (a) separates setup, warm-up, and production-SCF wall times for the four-component Dirac implementation (4c) and SWORD (S). Panel (b) reports the corresponding Dirac-to-SWORD wall-time ratio for the production SCF and for the complete calculation; values above unity indicate a lower wall time for SWORD.}
    \label{fig:timing_1el_v_2el}
\end{figure}

The timing comparison, however, reveals an important distinction from the one-electron case. Figure~\ref{fig:timing_1el_v_2el} shows that the reduction in propagation cost persists, but the overall gain is substantially reduced once the self-consistent potential must be rebuilt and applied. In the present implementation, operations outside the Green's-function propagation can become more expensive for the SWORD orbitals, offsetting part of the saving obtained from propagating fewer components. Consequently, the two-electron benchmarks already indicate that the practical performance of SWORD is determined not only by the nominal number of propagated components, but also by how the reconstructed functions are represented and subsequently processed by the adaptive MW machinery.

\subsection{Four-electron systems: non-local Hartree--Fock exchange}

The four-electron closed-shell case provides the smallest benchmark in the present hierarchy in which distinct occupied spatial orbitals generate both Coulomb and non-local exchange contributions. It therefore tests the method with all of the Dirac--Coulomb Hartree--Fock ingredients considered in this work while remaining sufficiently compact for detailed profiling. The Be-like sequence \ce{Be}, \ce{Ne^{6+}}, \ce{Ar^{14+}}, \ce{Kr^{32+}}, \ce{Xe^{50+}}, and \ce{Hg^{76+}} was examined. Kramers time-reversal symmetry was again exploited to reduce the number of explicitly treated occupied spinors\cite{KTRS}.


In contrast to the precision sweeps used for the one- and two-electron systems, the four-electron calculations were performed at MW7. At this stage the objective was to stress-test the formulation in the presence of non-local exchange and to examine its computational profile at high numerical precision. The many-electron propagation was implemented using Eq.~\eqref{eq:exact_N_el_Prop} without enforcing a strictly diagonal Fock matrix. Although this is not the least expensive realization of the algorithm, it provides a more general test of the coupling between occupied states and of its behavior during SCF convergence.

\begin{figure}[h]
    \centering
    \includegraphics[width=0.8\linewidth]{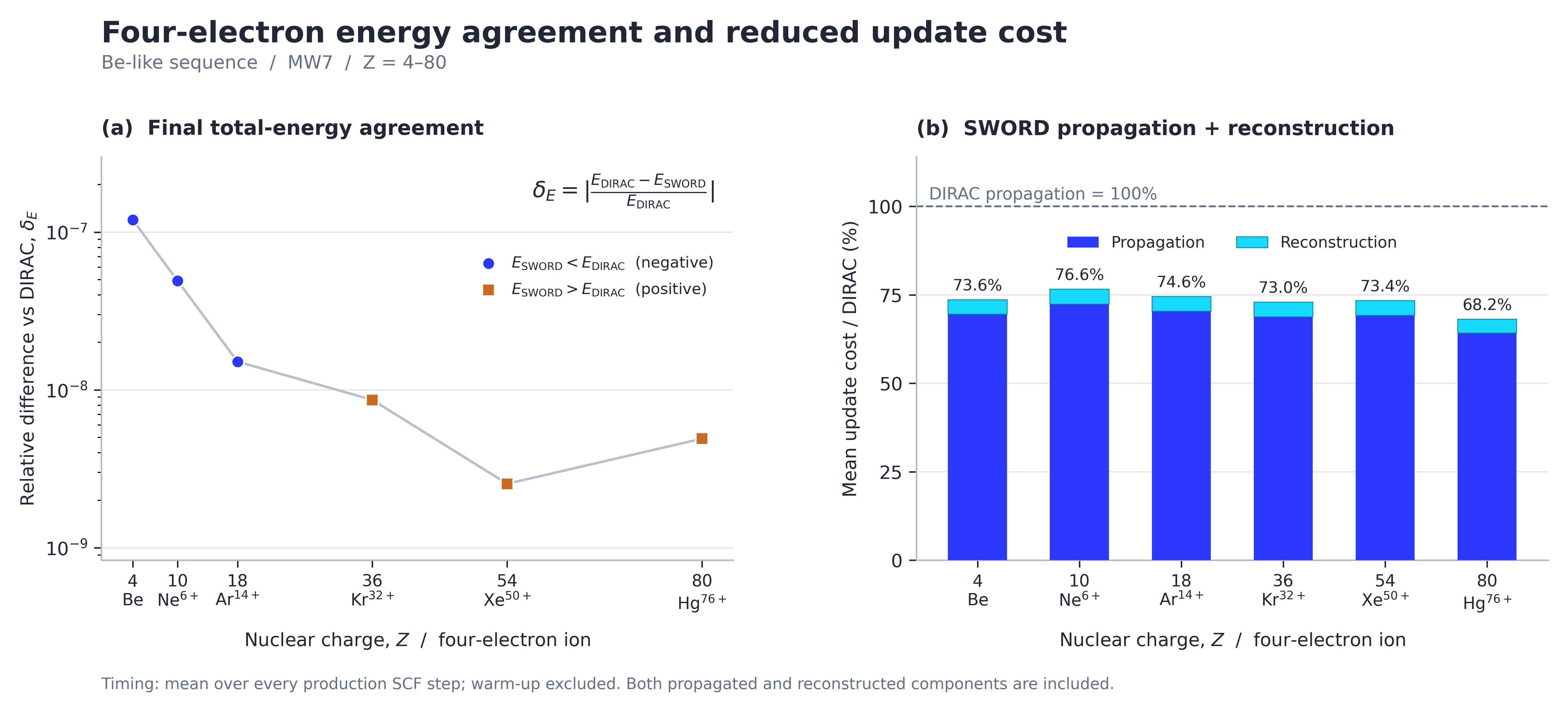}
    \caption{Four-electron Be-like sequence at MW7. (a) Relative difference between the final total energies obtained with the four-component Dirac implementation and SWORD. (b) Mean SWORD update cost relative to four-component propagation, separated into propagation and reconstruction contributions. The average is taken over the production SCF iterations; the warm-up stage is excluded.}
    \label{fig:energy_and_update_cost}
\end{figure}

\begin{figure}[h]
    \centering
    \includegraphics[width=0.8\linewidth]{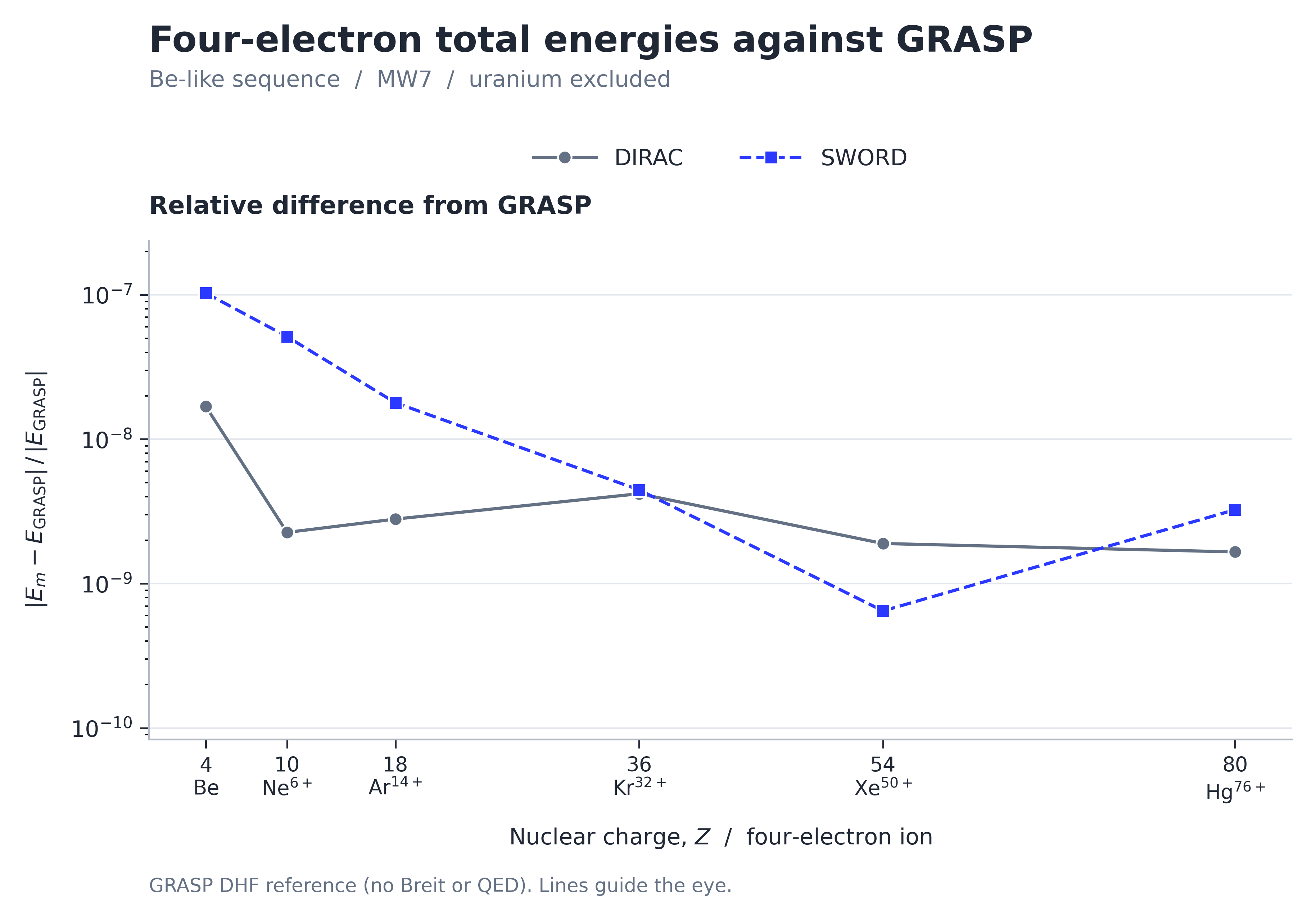}
    \caption{Relative total-energy error of the four-component Dirac and SWORD calculations with respect to \grasp{} Dirac--Hartree--Fock reference energies for the Be-like sequence at MW7. The same Dirac--Coulomb physical model is used for the comparison, without Breit or QED corrections.}
    \label{fig:gras_4l_ref}
\end{figure}

For every system considered, SWORD and the four-component implementation converged to total energies that agree within the numerical accuracy of the calculations. Figure~\ref{fig:energy_and_update_cost}(a) reports the relative difference
\begin{equation}
    \delta_E = \left|\frac{E_{\mathrm{Dirac}}-E_{\mathrm{SWORD}}}{E_{\mathrm{Dirac}}}\right|,
\end{equation}
which remains between approximately $10^{-7}$ and $10^{-9}$ over the Be-like sequence at MW7. The direct agreement between the two implementations is complemented by the independent \grasp{} comparison in Fig.~\ref{fig:gras_4l_ref}. Both methods reproduce the corresponding \grasp{} Dirac--Hartree--Fock energies to the expected numerical accuracy, showing that the agreement between SWORD and the four-component implementation is not merely an internal cancellation between two implementations sharing the same MW framework.

These results extend the one- and two-electron validation to a genuinely non-local Hartree--Fock problem. Within the precision investigated here, reconstruction of the second chirality remains compatible with the self-consistent Coulomb and exchange operators and does not introduce a detectable loss of final energy accuracy relative to the four-component propagation scheme.

\subsection{Computational performance of the four-electron calculations}

The four-electron calculations also make it possible to identify where the computational savings from reduced propagation are retained and where they are lost. Figure~\ref{fig:energy_and_update_cost}(b) shows that, including reconstruction, a SWORD orbital update requires on average approximately $73.3\%$ of the corresponding four-component propagation time over the systems examined. Reconstruction itself remains inexpensive; most of the SWORD update time is still associated with the Green's-function propagation. The reduction is therefore systematic, although it does not reach the idealized $50\%$ limit suggested by simply halving the number of explicitly propagated spinor components.

\begin{figure}[h]
    \centering
    \includegraphics[width=0.8\linewidth]{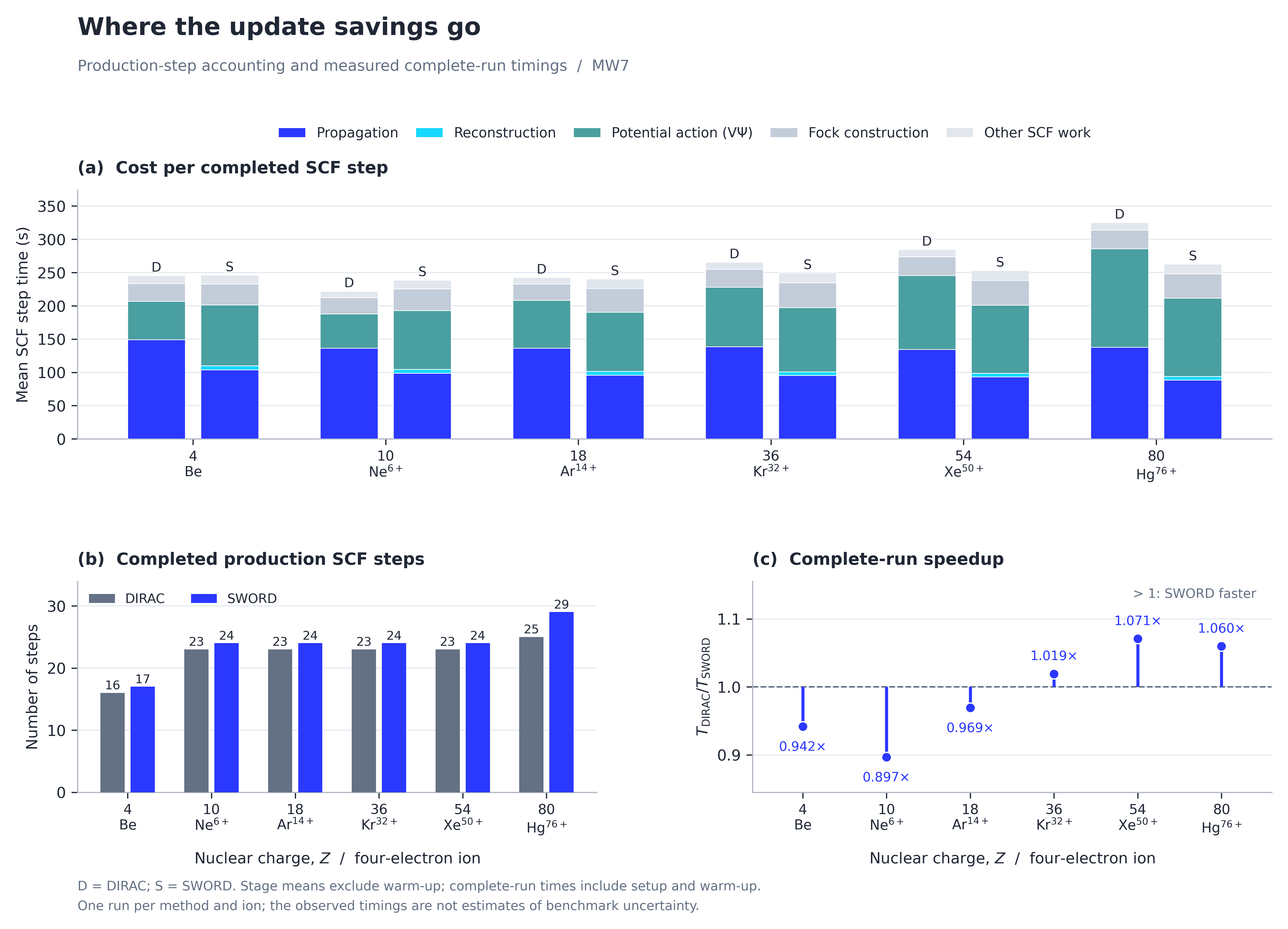}
    \caption{Runtime accounting for the four-electron Be-like sequence at MW7. (a) Mean cost per completed production SCF step, decomposed into propagation, reconstruction, potential action, Fock construction, and remaining SCF work. (b) Number of completed production SCF steps for the four-component Dirac implementation and SWORD. (c) Complete-run Dirac-to-SWORD wall-time ratio; values above unity indicate a lower wall time for SWORD. Stage averages exclude the warm-up, whereas complete-run timings include setup and warm-up.}
    \label{fig:runtime_accounting}
\end{figure}

A more complete accounting is shown in Fig.~\ref{fig:runtime_accounting}. The propagation stage is consistently cheaper with SWORD, but this saving does not translate directly into the same reduction in the total SCF-step cost. In particular, the potential action and Fock construction can be more expensive for the reconstructed SWORD orbitals. As a result, the measured complete-run speedup varies across the sequence: for some systems the present SWORD implementation is slightly slower overall, whereas for the heavier members it becomes slightly faster. The number of production SCF iterations remains very similar between the two formulations, with only a modest increase for SWORD and a somewhat larger difference for \ce{Hg^{76+}}. The timing differences therefore arise predominantly from the cost of the operations performed within each iteration rather than from qualitatively different convergence behavior.

The present profiling is particularly relevant because the implementation was intentionally developed as a proof of concept rather than as an optimized production code. SWORD currently reconstructs the full four-component spinor and then reuses routines designed for the four-component formulation. Consequently, only the propagation stage directly exploits the reduced representation, while subsequent operations do not yet benefit from an implementation tailored to the two-component structure. The measured wall times should therefore be interpreted as the performance of this first realization of SWORD, rather than as an optimized limit of the formulation.

The calculations further show that the reconstructed right-handed component can acquire a more highly refined MW representation than its four-component counterpart. Since the cost of subsequent adaptive operations depends on the representation of the functions on which they act, this additional refinement provides a plausible mechanism for part of the increased cost observed in the potential and Fock stages. In the present implementation, this effect can therefore erode a fraction of the saving achieved during propagation even though the reconstruction itself is inexpensive.

%

\subsection{Numerical differentiation and SCF warm-up}\label{sec:warmup_considerations}

The proof-of-concept calculations also revealed a numerical sensitivity associated with the construction and repeated use of differentiated MW functions. This behavior was most evident for heavy elements when the raw kinetically balanced guess of Eq.~\eqref{eq:weyl_balanced_guess} was supplied directly to the high-precision SWORD iteration. Under these conditions, the first SCF iterations could develop fine-scale numerical structure that triggered excessive adaptive refinement and, in the most severe cases, prevented a stable start of the SCF procedure.

Differential operators require particular care in an adaptive MW representation because differentiation emphasizes short-wavelength components of a numerical function. Small-scale artifacts that are negligible in the orbital itself can therefore become more visible after differentiation and can trigger additional local refinement\cite{Yanai2004_derivatives,Anderson2019-bx_Derivatives}. This is especially relevant to SWORD because the opposite chirality is reconstructed through an expression containing $\hat{\pi}\psi_L$, so numerical structure present in the propagated left-handed component is directly exposed to a momentum operator during every reconstruction.


The sensitivity becomes more pronounced for increasing nuclear charge because core orbitals become more strongly localized and contain larger momentum components. Correspondingly, the momentum-dependent correction in the kinetically balanced Weyl guess,
\begin{equation}
    \frac{\hat{\pi}}{2mc}\Phi,
\end{equation}
becomes increasingly important relative to the leading non-relativistic component. Inspection of the one-electron calculations showed that small, rapidly varying features could remain in the SWORD orbitals after the first updates and become more apparent after application of $\hat{\pi}$, whereas the corresponding four-component propagation produced smoother starting orbitals under the same high-precision workflow.


A simple warm-up procedure was sufficient to remove this sensitivity in the calculations reported here. The initial spinor was first propagated for a few iterations at four-component level and deliberately low numerical precision. This suppresses fine-scale features that are irrelevant at the accuracy of the starting guess before the orbitals are supplied to either production algorithm. The same smoothed orbitals were then used to initialize both the four-component and SWORD calculations, preserving a common starting point for the subsequent comparison.

This behavior is consistent with the dynamic-precision strategy commonly used in MW SCF calculations: early iterations need not be carried out at the final target precision, and the numerical representation can be tightened as the orbitals approach self-consistency. In the present work the warm-up serves as a minimal substitute for such production-level machinery, which was not implemented in the prototype. The observed sensitivity should therefore be viewed primarily as an implementation issue associated with starting a derivative-sensitive adaptive calculation directly at high precision, rather than as an intrinsic limitation of the SWORD formulation.


Overall, the benchmark hierarchy provides three complementary tests of the formulation. The one-electron calculations validate the reduced propagation and reconstruction in a fixed external potential; the two-electron calculations show that the same procedure remains accurate when the local potential is determined self-consistently; and the four-electron calculations extend the validation to non-local Hartree--Fock exchange. Across all three levels, SWORD reproduces the four-component energies within the requested numerical accuracy while reducing the cost of the propagation stage. The remaining gap between this reduction and the total wall-clock performance is concentrated in operations that have not yet been reformulated or optimized for the reduced representation, leaving substantial scope for a dedicated SWORD implementation to exploit the computational advantage of propagating only one chirality.

\section{Conclusions}

In this work, we introduced SWORD, a two-component reformulation of the Dirac equation based on the Weyl representation and developed within an adaptive multiwavelet framework. The Weyl representation provides a natural setting for this approach. By changing the representation of the matrices entering the Dirac equation, the differential operators are moved to the diagonal blocks of the Hamiltonian, while the off-diagonal coupling reduces to the scalar mass term. This structure allows a simple and exact coupling relation to be established between the two chiral components of the Dirac spinor.

The central idea of SWORD is to exploit this coupling directly: only one chirality is propagated explicitly, while the other is reconstructed self-consistently through the exact coupling relation. In this way, the original four-component Dirac problem is not replaced by an approximate two-component Hamiltonian. Instead, it is reorganized such that only half of the spinorial degrees of freedom undergo the computationally expensive Green’s-function propagation.

A key consequence of this reformulation is that the resulting propagation equation takes the form of an inhomogeneous Helmholtz equation for the propagated chirality, which can be inverted naturally through the application of the corresponding Green’s function. This makes multiwavelets particularly well suited to the present formulation. Their role, however, goes beyond providing an efficient machinery for differential operators and Green’s-function convolutions: their adaptive and systematically improvable representation also provides the flexibility required to describe the chiral spinor components without imposing a predefined analytic form on the basis functions.

The formulation was tested through a sequence of increasingly complete problems. One-electron calculations first established the behavior of the reduced propagation and reconstruction in a fixed external potential. Two-electron closed-shell systems then introduced a self-consistent Coulomb potential, while the four-electron calculations provided the first tests in the presence of non-local Hartree--Fock exchange. Across all systems considered, SWORD reproduced the corresponding four-component results within the requested numerical precision, with both formulations also showing consistent agreement with the independent \grasp{} reference calculations. The calculations therefore provide a proof of concept that the reduced Weyl propagation can be used without sacrificing the four-component Dirac--Coulomb energetics investigated in this work.

The numerical results also show where the present formulation is already advantageous and where further development is required. Propagating only one chirality consistently reduces the cost of the Green's-function update, while reconstruction of the complementary component represents only a small fraction of this cost. In the four-electron calculations, the complete SWORD update, including reconstruction, required on average approximately $73.3\%$ of the corresponding four-component propagation time. This is a substantial reduction, although it remains above the idealized $50\%$ limit suggested by the number of explicitly propagated components.

Importantly, the same reduction is not yet obtained for the total SCF wall time. The present implementation was designed as a proof of concept and deliberately reuses much of the existing four-component machinery after reconstructing the full spinor. Consequently, operations such as the evaluation of the potential and construction of the Fock matrix do not yet exploit the reduced representation and can become more expensive for the reconstructed orbitals. The additional adaptive refinement observed in the reconstructed component provides one possible contribution to this overhead and deserves further investigation. The timings reported here should therefore be interpreted as the performance of the present prototype, rather than as the computational limit of the SWORD formulation.

The calculations also exposed numerical aspects that will be important in a production implementation. In particular, the repeated use of differential operators makes the reduced formulation sensitive to fine-scale numerical structure in a raw starting guess, especially for heavy nuclei. In the present implementation this was addressed by a low-precision warm-up of the initial orbitals before the production SCF calculation. More generally, these observations point to the importance of dynamic precision control and of a representation specifically adapted to the reconstructed component when SWORD is implemented in a complete multiwavelet electronic-structure code.

Several natural extensions remain open. A first objective is to reformulate as much as possible of the effective-potential action directly in terms of the propagated chirality, avoiding the reconstruction and subsequent four-component treatment whenever it is not required. The relation between numerical precision, adaptive refinement, and reconstruction should also be characterized more systematically, with the aim of preventing unnecessary fine-scale structure and bringing the measured propagation cost closer to the reduction suggested by the formalism. Other possible developments to be explored concern reformulating the whole algorithm without ever needing to address or calculate explicitly the right component. Moreover, given the tight precision control of multiwaveles, it would possible to explore unique precision assessment criteria for convergence, as well as well-tuned numerically exact approximations, avoiding to calculate some heavy terms if the expected norm contribution fall below the precision threshold.  

Beyond the Dirac--Coulomb Hartree--Fock model considered here, the same framework can be investigated for density-functional methods and relativistic two-electron interactions such as the Breit contribution.

A natural long-term step is therefore to move beyond the present \vampyr{} prototype and incorporate the method into a production multiwavelet environment such as \mrchem{}, where established SCF infrastructure, dynamic precision control, and optimized low-level operations could be combined with a SWORD-specific implementation. Such an implementation would make it possible to determine how much of the formal reduction in propagation cost survives for larger and chemically relevant systems, and whether further parts of the relativistic calculation can be expressed directly within the reduced Weyl representation.

The present results do not establish SWORD as an optimized alternative to existing relativistic methods; they establish that the underlying computational idea works. The Weyl representation provides an exact relation between the two chiral sectors that can be exploited algorithmically, and the resulting method reproduces four-component Dirac--Hartree--Fock energies while explicitly propagating only one of them. The remaining performance limitations arise in parts of the prototype that have not yet been redesigned around this reduced structure. This leaves considerable room for further development and, more broadly, suggests that the choice of representation in relativistic quantum chemistry can be used not only to reorganize the equations, but also as a practical principle for designing new numerical algorithms.

\section{Acknowledgments}
We acknowledge support from the Research Council of Norway through its Centres of Excellence scheme (262695), as well as through the FRIPRO grant ReMRChem (324590).

\section*{Declarations}

We use the CRediT taxonomy of contributor roles.\cite{Allen2014-rd,Brand2015-qc}
The ``Investigation'' role also includes the ``Methodology'', ``Software'', and ``Validation'' roles.
The ``Analysis'' role also includes the ``Formal analysis'' and ``Visualization''
roles. The ``Funding acquisition'' role also includes the ``Resources'' role.
We visualize contributor roles in the following authorship attribution matrix,

\begin{table}[ht]
\caption{Levels of contribution: \textcolor{blue!100}{major}, \textcolor{blue!25}{minor}.}
\label{tbl:contribs}
\begin{tabular}{lcccc}
\hline\hline
                              &  JM    & QP      & LF     \\ \hline
    Conceptualization         & \major &   \minor &                 \\ 
Investigation             & \major &   & \minor & \\ 
    Data curation             & \major &    & \minor \\ 
    Supervision               &  \major      &    & \minor \\ 
    Writing -- original draft & \major & \minor  & \minor \\
    Writing -- revisions      & \major &  \minor  & \major \\
    Funding acquisition       &        &         & \major \\ 
    Project administration    &        &   & \major \\ \hline\hline
\end{tabular}
\end{table}

\bibliography{references}
\end{document}